\documentclass[twocolumn,aps,prd,amsmath,amssymb,preprintnumbers,nofootinbib,superscriptaddress]{revtex4-1}
\usepackage{graphicx,caption}
\usepackage{subcaption}
\usepackage{soul}
\usepackage{hyperref,url}
\hypersetup{
	colorlinks=true,       
	linkcolor=blue,          
	citecolor=blue,        
}
\usepackage{comment}

\begin{document}
\setcounter{footnote}{0}

\title{Covariant LTB collapse in models of loop quantum gravity}

\author{Martin Bojowald}
\email{bojowald@psu.edu}
\affiliation{Institute for Gravitation and the Cosmos,
The Pennsylvania State University, 104 Davey Lab, University Park, PA 16802, USA}

\author{Erick I.\ Duque}
\email{eqd5272@psu.edu}
\affiliation{Institute for Gravitation and the Cosmos,
The Pennsylvania State University, 104 Davey Lab, University Park, PA 16802, USA}

\author{Dennis Hartmann}
\email{dennis.hartmann@psu.edu}
\affiliation{Institute for Gravitation and the Cosmos,
The Pennsylvania State University, 104 Davey Lab, University Park, PA 16802, USA}
\affiliation{Department of Astronomy and Astrophysics, The Pennsylvania State University, 525 Davey Lab, University Park, PA 16802, USA}

\begin{abstract}
  Models of gravitational collapse provide important means to test whether
  non-classical space-time effects motivated for instance by quantum gravity
  can be realized in generic ways in physically relevant situations. Here, a
  detailed analysis of marginally bound Lema\^{\i}tre--Tolman--Bondi
  space-times is given in emergent modified gravity, which in particular
  includes a covariant formulation of holonomy modifications usually
  considered in models of loop quantum gravity. As a result, generic collapse
  in this setting is shown to imply a physical singularity that removes the
  bouncing behavior seen in vacuum space-times with the same type of
  modifications.
\end{abstract}

\maketitle

\section{Introduction}

Black holes present a fascinating problem in general relativity and possible
extensions such as modified or quantum gravity. Being described by a small
parameter space, they appear as rather simple objects from a classical
perspective. However, it is often postulated that quantum effects should start
to dominate near the singularity of a black hole, where curvature is large and
divergences lead to undefined quantities in the classical theory. Various
approaches to quantum gravity have therefore suggested modifications of the
classical behavior of space-time with the main goal of resolving physical
singularities.

Unraveling the mysteries of black holes requires an understanding of their
formation. The conditions and mechanisms that cause the collapse of matter to
a black hole are numerous. In this paper, we shall focus on the
Lema\^{\i}tre--Tolman--Bondi (LTB) collapse model of dust matter within
emergent modified gravity, extending the original analysis with canonical
methods given in \cite{LTB,LTBII}. We leave the critical collapse of scalar
fields to a future paper.  Both types of collapse models are well-established
and have been studied from different perspectives. LTB collapse is of interest
in the present paper in order to study a manifestly covariant model of matter
collapse in which partial analytical results are available.

Emergent modified gravity \cite{Higher,HigherCov} is an approach to modified
gravity crucially based on the canonical formulation of general
relativity. Because canonical expressions do not require space-time integrals,
they impose weaker assumptions on the basic setup than space-time scalars
required for effective space-time actions. As a result, new theories of
modified gravity are available, at least in spherical symmetry and polarized
Gowdy systems \cite{EmergentGowdy}, even if one works at the classical
derivative order of equations of motion and does not include higher time
derivatives or extra fields. The space-time geometries resulting from
solutions of the theory are Riemannian, but the line element is not directly
given by one of the fundamental fields and instead has to be derived from the
field equations. Couplings to perfect fluids, scalar matter, and the
electromagnetic field have been derived in
\cite{EmergentFluid,EmergentScalar,EmergentEM}.

Prior studies of black holes in emergent modified gravity or related
approaches \cite{SphSymmMinCoup} have revealed interesting results, such as
the potential for non-singular behavior
\cite{SphSymmEff,SphSymmEff2,SphSymmEff3} or, for a special set of modifications, dynamical signature change of metric components at classical
scales \cite{EmergentSig} which demonstrates the non-standard nature of new
space-time solutions. Collapse models have been analyzed in an
Oppenheimer--Snyder version \cite{EmergentFluid}, and already in LTB models
\cite{SphSymmLTB1,SphSymmLTB2}.
The latter models were restricted to a subset
of modification functions, in particular with a constant holonomy length, that are known to lead to large deviations from classical behavior even on small curvature scales.
One of the main aims of our work is to provide a sufficiently general formulation and analysis of LTB models that also includes decreasing holonomy lengths.

Emergent modified gravity includes modification functions that clarify the
appearance of holonomy modifications of loop quantum gravity
\cite{EmergentMubar}. It therefore provides a unique systematic way of turning
preliminary constructions that implement loop effects in a fixed gauge or in a
deparameterized theory into fully covariant space-time models. (Alternative
constructions of a limited set of such models have been presented in
\cite{MassCovariance}. However, as discussed in \cite{Lessons}, they are
special cases of emergent modified gravity and do not constitute a systematic
derivation of modified gravity within a reliable setting of effective field
theory.) In particular, emergent modified gravity can be used to identify
problems in LTB models of loop quantum gravity, such as non-covariance in the
recent \cite{LTBShock,LoopGauge2,LoopShock,LTBPol,LTBPolII}, and amend them by
including symmetry-restoring terms in the Hamiltonian constraint.
It is important to note that emergent modified gravity is not
  a quantum theory, though some modifications may be motivated by it, in the
  present case of interest by loop quantum gravity (LQG). However, the theory is non-classical in the sense that it is inequivalent to general relativity. In the following, we use the term \emph{classical} to refer to general relativity. In particular, the classical limit of emergent modified gravity is defined as general relativity.

An issue that has been discussed extensively is the physical meaning of
shell-crossing singularities and potential shock waves that may occur at a
turning point of collapse, a possible cure of the classical curvature
singularity. A careful analysis has been given in \cite{LoopShock} where it
was found important to compare results from different choices of coordinates
and space-time slicings. However, the same paper pointed out that available
models had not been shown to be covariant, which would be an important
prerequisite to any physical analysis of space-time effects. Here, we complete
the discussion by constructing a corresponding formulation that is explicitly
shown to be covariant. Our main result is surprising: While covariance can be
restored, we find that LTB models with modifications motivated by loop quantum
gravity have a physical, rather than shell-crossing, singularity even though
corresponding vacuum models are generically non-singular. A similar result
with dust \cite{EmergentFluid} and scalar matter \cite{EmergentScalar} had
already been found in emergent modified gravity, and
\cite{SphSymmLTB1,SphSymmLTB2} extended this effect to LTB collapse (with
constant holonomy length). The latter papers then introduced a
conformal transformation to a non-singular solution and argued that this
should be the physical one. A second aim of the present paper is to critically
examine this proposal, which we find inadmissible when seen from the
perspective of a modified theory rather than a single modified solution.

The result of a generic re-appearance of physical singularities in LTB
collapse raises important questions about the reliability of singularity
resolution or other effects found in simple models of loop quantum gravity.
It highlights the importance of constructing fully covariant space-time models
whenever a space-time analysis based on effective line elements is performed:
The singular behavior can be traced back to a crucial factor in the space-time
metric that had been missed in non-covariant models.  It is also important to
include a generic set of modification functions in one's class of theories,
such that all potential effects at the same order of derivatives can be
included. Emergent modified gravity fulfills this condition from the point of
view of effective field theory.

Our construction of LTB models in emergent modified gravity is presented in
Section~\ref{s:LTB}. We pay special attention to the non-fundamental nature of
LTB conditions, which are motivated by convenient forms of solutions rather
than fundamental principles such as symmetries. The procedures for choosing
and implementing such conditions are therefore not as unique as, say, symmetry
reduction. Moreover, the classical conditions are motivated by a specific form
of the metric directly constructed from kinematical phase-space variables such
as the spatial metric or triad, for which equations of motion are first
formulated and then solved. Models of emergent modified gravity, and therefore
covariant models of loop quantum gravity, are crucially different because
their space-time metric in terms of phase-space variables is emergent and not
known before equations of motion are imposed.

Our detailed discussion of these issues is followed in
Section~\ref{s:Collapse} by an analysis of various collapse solutions.  In
particular, in Section~\ref{s:Generic} we demonstrate that local covariant
marginal LTB solutions with holonomy modifications are strictly reflection
symmetric around a hypersurface of maximum curvature. This symmetry, which is
explicitly demonstrated by specific solutions in Sections~\ref{s:Constant} and
\ref{s:Scale}, implies that shock waves cannot form near maximum curvature
unless they had already been present in the initial data of collapse.  At the
same time, we will see that the maximum curvature is infinite, implying a
physical singularity even in the presence of modifications from loop quantum
gravity. Our concluding Section~\ref{s:Concl} emphasizes several key
advantages of embedding models of loop quantum gravity within emergent
modified gravity that were crucial for our analysis.

\section{LTB models in emergent modified gravity}
\label{s:LTB}

Classical line elements of LTB type \cite{LTBReview} are given by
\begin{equation}\label{eq:LTB metric}
    {\rm d} s^2 = - {\rm d} t^2 + \frac{(R')^2}{1+\kappa (x)} {\rm d} x^2 +
    R^2  ({\rm d}\vartheta^2+\sin^2\vartheta{\rm d}\varphi^2)
  \end{equation}
  where $\kappa(x)$ is a non-dynamical radial function that distinguishes
  different models. The dynamical variable, $R(t,x)$, is a function of
  the time and radial coordinates subject to the differential equation
  \begin{equation}
    \dot{R}^2 = \kappa (x) + \frac{2 m(x)}{R}
  \end{equation}
  with another radial function, $m(x)$, that is determined by the initial mass
  distribution: The energy density of matter is given by
  \begin{equation}
    T_{tt} = \frac{m'}{4 \pi G R^2 R'}
    \,.
  \end{equation}
  Appendix~\ref{a:LTB} provides a more detailed review as well as several examples.

\subsection{LTB models in canonical gravity}
  
  For an implementation in emergent modified gravity, the canonical
  formulation of an LTB reduction is crucial. Since LTB models are spherically
  symmetric, they can be expressed within the general class of line elements
  of the form
  \begin{equation}\label{eq:ADM line element - spherical - GR}
    {\rm d} s^2 = - N^2 {\rm d} t^2 + \frac{(E^\varphi)^2}{E^x} ( {\rm d} x +
    N^x {\rm d} t )^2 + E^x ({\rm d} \vartheta^2+\sin^2\vartheta{\rm d}\varphi^2)
  \end{equation}
  with free functions $N(t,x)$, $N^x(t,x)$, $E^x(t,x)$ and
  $E^{\varphi}(t,x)$. In this form, $N$ and $N^x$ are the usual lapse function
  and radial shift vector, while $E^x$ and $E^{\varphi}$ are components of a
  densitized triad that may be used to parameterize spatial metric
  components. The orientation of the triad is given by the sign of $E^x$,
  which we will assume to be positive throughout this paper.

\subsubsection{Vacuum properties}
  
  These variables are convenient in a canonical formulation, where, together
  with their momenta $K_x$ and $K_{\varphi}$, they are
  subject to the Hamiltonian constraint \cite{SphSymm,SphSymmHam}
\begin{widetext}
  \begin{equation}
    H^{\rm grav} [N] = \int {\rm d} x\ N \left[ \frac{((E^x)')^2}{8 \sqrt{|E^x|} E^\varphi}
    - \frac{E^\varphi}{2 \sqrt{|E^x|}}
    - \frac{E^\varphi K_\varphi^2}{2 \sqrt{|E^x|}}
    - 2 K_\varphi \sqrt{|E^x|} K_x
    - \frac{\sqrt{|E^x|} (E^x)' (E^\varphi)'}{2 (E^\varphi)^2} 
    + \frac{\sqrt{|E^x|} (E^x)''}{2 E^\varphi}
    \right] =0
    \label{eq:Hamiltonian constraint - spherical symmetry - Gauss solved}
\end{equation}
\end{widetext}
and the diffeomorphism constraint
\begin{eqnarray}
    H_x^{\rm grav} [N^x] &=& \int {\rm d} x\ N^r \left( K_\varphi' E^\varphi - K_x (E^x)' \right)=0
    \label{eq:Diffeomorphism constraint - spherical symmetry- Gauss solved}
\end{eqnarray}
in vacuum space-times. The LTB form of the line element is obtained with the
gauge-fixing condition
\begin{equation}\label{eq:LTB gauge}
  N=1\quad,\quad N^x=0
\end{equation}
and the LTB condition
\begin{equation} \label{eq:LTB condition}
    E^\varphi = \frac{(E^x)'}{2 \sqrt{1+\kappa (x)}} \,.
\end{equation}
The dynamical function $R(t,x)$ is then identified with $R=\sqrt{E^x}$.

In a complete LTB solution, the LTB condition 
\begin{equation}\label{eq:LTB constraint - GR}
    L_1 \equiv E^\varphi - \frac{(E^x)'}{2 \sqrt{1+\kappa (x)}}=0
  \end{equation}
  must be fulfilled at all times, and therefore $L_1$ must Poisson commute
  with the Hamiltonian $H[N]+H_x[N^x]=H[1]$ on-shell, when the constraints and
  equations of motion hold \cite{LTB,LTBII}. As shown in more detail in
  App.~\ref{a:LTB}, this condition evaluates to
\begin{eqnarray}
    \{L_1,H[1]\} &=&
    \left( \frac{K_\varphi}{\sqrt{|E^x|}}
    + \frac{\sqrt{|E^x|}}{E^\varphi} 2 K_x \right) L_1
    \nonumber\\
    &&\quad
    - \frac{\sqrt{|E^x|}}{E^\varphi} \frac{H_x^{\rm grav}}{\sqrt{1+\kappa (x)}}
  \end{eqnarray}
  and therefore does vanish on-shell for vacuum LTB solutions, where
  $H_x^{\rm grav}=0$.

  For further applications in the modified case, we note that our minimal
  implementation of the LTB condition, following \cite{LTB,LTBII} but in
  contrast to \cite{LTBPol}, is suitable for a phenomenological
  reduction. Compared with symmetry reduction, for instance from the full
  theory to spherical symmetry, the LTB condition does not have a fundamental
  interpretation but rather imposes a restriction to a convenient but not
  physically characterized set of solutions. Such a phenomenological reduction
  is distinguished from a fundamental reduction in that the resulting solution
  space need not be equipped with a phase-space structure. Thus, we have
  minimally required that the LTB condition $L_1$ is preserved in time, which
  led us to an on-shell condition that is automatically fulfilled for the
  classical dynamics. Since LTB models only implement a specific solutions
  space, an on-shell treatment is sufficient. It is not necessary to analyze
  our $L_2$ off-shell and determine, for instance, whether the pair
  $(L_1,L_2)$ is first class or second class and, in the latter case, derive
  the corresponding Dirac bracket. For phenomenological reductions, it is
  possible to implement a single condition, $L_1=0$. In a fundamental
  reduction, a single condition would always be first class and requires
  factoring out gauge transformations in order to obtain an even-dimensional
  phase space with a non-degenerate symplectic structure. A phenomenological
  reduction, by contrast, may easily work with an odd-dimensional solution
  space since no reduced phase-space structure is required. It is important to
  remember the distinction between phenomenological and fundamental reduction
  when the LTB condition is evaluated in modified gravity.

  In what follows, we demonstrate that standard LTB solutions are obtained
  with this procedure when applied to general relativity. Any implementation
  of the same procedure in modified gravity in canonical form therefore
  provides valid modifications of the LTB solution space. It is, however,
  consistent to follow a more involved process such as the full implementation
  of the formal Dirac algorithm as performed in \cite{LTBPol}. The
  availability of different methods for phenomenological reduction constitutes
  a source of ambiguities in models of modified gravity.  For this reason, LTB
  models, in contrast, for instance, to models of scalar collapse in spherical
  symmetry, do not provide conclusive answers to fundamental questions such as
  the fate of black-hole singularities or potential observational
  implications. Nevertheless, they play an important role as tractable testing
  grounds of various ideas in modified and quantum gravity.

\subsubsection{Dust}
  
Classical LTB models are compatible with dust-like matter as the source of the
mass function $m(x)$. As an example of perfect fluids, dust is canonically
described by two spatial scalars, $T(x)$ and $X(x)$, that can be viewed as
space-time coordinates of evolving fluid particles; see also
\cite{BrownKuchar}. Their momenta $P_T$ and $P_X$ are spatial densities and
represent energy and momentum of the particles, which therefore have the
radial velocity $P_X/P_T$. They appear in the matter contributions to the
Hamiltonian constraint,
\begin{equation}
    H^{\rm matter}_s =
    \sqrt{ s P_T^2 + \frac{E^x}{(E^\varphi)^2} (H_x^{\rm matter})^2}
    - \sqrt{\det{q}}\; P \left(\frac{P_X^2}{P_T^2}\right),
    \label{eq:Hamiltonian constraint - PF - spherical - classical}
\end{equation}
and the diffeomorphism constraint,
\begin{equation}
    H_x^{\rm matter} =
    P_T T'
    + P_X X'\,.
    \label{eq:Diffeomorphism constraint - PF - spherical}
  \end{equation}
In the first expression, the constant $s$ takes the value $s=1$ for massive
particles, and $s=0$ for massless ones. The pressure function $P(P_X^2/P_T^2)$
vanishes in the case of dust, and $\sqrt{\det q}=\sqrt{E^x}E^{\varphi}$ is the
spatial volume element.

For dust, we have to evaluate the preservation of the LTB condition $L_1$ with
the Hamiltonian $H[1]=H^{\rm grav}[1]+H^{\rm matter}[1]$, which leads to
\begin{equation}
  \{L_1,H[1]\}=\frac{\sqrt{E^x}}{E^{\varphi}} \frac{H_x^{\rm
      matter}}{\sqrt{1+\kappa(x)}}
\end{equation}
on-shell. Stability of the LTB condition therefore requires $H_x^{\rm matter}=0$ in the LTB gauge or slicing, which implies that the frame is co-moving with the fluid. We then fix $s=1$ because a physical
co-moving frame exists only if the fluid particles are massive.

\subsection{Emergent modified gravity}

The spherically symmetric constraints have Poisson brackets
\begin{eqnarray}
    \!\!\{ H_x [N^x] , H_x[M^x] \} \!\!&=&\!\! - H_x [M^x (N^x)'-N^x (M^x)']
    \label{H_x,H_x bracket}
    \\
    \!\!\{ H [N] , H_x [M^x] \} \!\!&=&\!\! - H[M^x N'] 
    \label{eq:H,H_x bracket - classical}
    \\
    \!\!\{ H [N] , H[M] \} \!\!&=&\!\! - H_x \left[ q^{x x} \left( M N' - N M' \right)\right]
    \label{eq:H,H bracket - classical}
\end{eqnarray}
with the inverse radial metric $q^{xx}=E^x/(E^{\varphi})^2$. They
characterize canonical gauge transformations as hypersurface deformations. The
gauge transformations of spatial metric components (or the densitized triad) and their momenta, or any
phase-space function depending on them, are
directly obtained in canonical form as
\begin{equation} \label{deltaf}
  \delta_{\epsilon}f(E,K) = \{f(E,K),H[\epsilon^0]+H_x[\epsilon^x]\}
\end{equation}
with two components $\epsilon^0$ and $\epsilon^x$ of the gauge function
$\epsilon$.

Gauge transformations of lapse and shift are more indirect because these
functions do not have momenta in canonical gravity, but as metric components
should nevertheless change when a spatial hypersurface is deformed in a new
slicing. Their transformations,
\begin{eqnarray}
    \delta_\epsilon N &=& \dot{\epsilon}^0 + \epsilon^x N' - N^x (\epsilon^0)' \\
    \delta_\epsilon N^x &=& \dot{\epsilon}^x + \epsilon^x (N^x)' - N^x (\epsilon^x)'
    \nonumber\\
    &&\quad+ q^{x x} (\epsilon^0 N' - N (\epsilon^0)' )
    \label{eq:Off-shell gauge transformations for lapse and shift - spherical - classical}
\end{eqnarray}
follow from an implementation of consistency of evolution generated by
$H[N]+H_x[N^x]$ with gauge transformations (\ref{deltaf}) generated by the same constraints
\cite{CUP,HypDef}. In particular, the structure function $q^{xx}$ in (\ref{eq:H,H
  bracket - classical}) appears in the transformations (\ref{eq:Off-shell gauge
  transformations for lapse and shift - spherical - classical}).

The canonical theory is covariant because of a crucial on-shell equivalence of
gauge transformations with space-time Lie derivatives of the metric, expressed
as
\begin{eqnarray}
    \delta_\epsilon g_{\mu \nu} \big|_{{\rm O.S.}} &=&
    \mathcal{L}_{\xi} g_{\mu \nu} \big|_{{\rm O.S.}}
    \,.
    \label{eq:Covariance condition - classical}
\end{eqnarray}
Here, ``O.S.'' refers to on-shell, requiring an evaluation of the condition on
the solution space of the theory. In this way, it is possible to relate
momenta in the canonical expressions on the left with time components of the
Lie derivative on the right. Moreover, the space-time vector field $\xi^{\mu}$
is related to the gauge functions by
\begin{equation}
  \xi^\mu = \epsilon^0 n^\mu + \epsilon^x s^\mu = \xi^t t^\mu + \xi^x s^\mu\,,
\end{equation}
transforming the normal frame used for $\epsilon$ to the coordinate frame used
for $\xi$. These two frames are related by the expression
$t^{\mu}=N n^{\mu}+N^x s^{\mu}$ for the time-evolution vector field where
$n^\mu$ is the unit normal to a spatial hypersurface and a $s^{\mu}$ a radial
vector field tangential to the hypersurface. The components are
therefore related by
\begin{equation}
    \xi^t = \frac{\epsilon^0}{N}
    \quad , \quad
    \xi^x = \epsilon^x - \frac{\epsilon^0}{N} N^x
    \,.
    \label{eq:Diffeomorphism generator projection - spherical}
\end{equation}

This classical structure of space-time can be generalized if the original
Hamiltonian constraint is modified such that the hypersurface-deformation
brackets and the covariance condition are preserved. A theory of emergent
modified gravity is then defined by a Hamiltonian constraint $\tilde{H}$ such that
\begin{eqnarray}
    \!\!\{ H_x [N^x] , H_x[M^x] \} \!\!&=&\!\! - H_x [M^x (N^x)'-N^x (M^x)']
    \label{H_x,H_x bracket - emergent}
\\
    \!\!\{ \tilde{H} [N] , H_x [M^x] \} \!\!&=&\!\! - \tilde{H}[M^x N'] 
    \ ,
    \label{eq:H,H_x bracket - emergent}
    \\
    \!\!\{ \tilde{H} [N] , \tilde{H}[M] \} \!\!&=&\!\! - H_x \left[ \tilde{q}^{x x} \left( M N' - N M'\right)\right]
    \label{eq:H,H bracket - emergent}
\end{eqnarray}
with a phase-space function $\tilde{q}^{xx}$ for which
\begin{eqnarray}
    \delta_\epsilon \tilde{g}_{\mu \nu} \big|_{{\rm O.S.}} &=&
    \mathcal{L}_{\xi} \tilde{g}_{\mu \nu} \big|_{{\rm O.S.}}
    \label{eq:Covariance condition - emergent}
\end{eqnarray}
holds with a space-time metric $\tilde{g}_{\mu\nu}$ derived from the emergent line element
\begin{equation}
    {\rm d} s^2 = - N^2 {\rm d} t^2 + \tilde{q}_{xx} ({\rm d} x + N^x {\rm d} t)^2
    + E^x {\rm d} \Omega^2
    \label{eq:ADM line element - emergent - spherical}
\end{equation}
that makes use of the inverse $\tilde{q}_{xx}$ of the new structure function
$\tilde{q}^{xx}$. In contrast to standard general relativity,
$\tilde{q}^{xx}$ need not be defined a priori in terms of the fundamental
fields. It must be derived from hypersurface-deformation brackets and the
covariance condition and is therefore emergent. As the new structure function,
the emergent spatial metric
then appears in the gauge transformations
\begin{eqnarray}
    \delta_\epsilon N &=& \dot{\epsilon}^0 + \epsilon^x N' - N^x (\epsilon^0)' ,\\
    \delta_\epsilon N^x &=& \dot{\epsilon}^x + \epsilon^x (N^x)' - N^x (\epsilon^x)'
    \nonumber\\
    &&\quad
    + \tilde{q}^{x x} \left(\epsilon^0 N' - N (\epsilon^0)' \right)
    \label{eq:Off-shell gauge transformations for lapse and shift - spherical - emergent}
\end{eqnarray}
of lapse and shift.

Hamiltonian constraints compatible with these conditions have been evaluated
systematically in a derivative expansion that stops at the classical order of
at most two spatial derivatives in each term. Even with this classical order,
non-trivial modifications are possible because emergence of the metric is a
new property that turns out to relax several implicit assumptions in
action-based constructions of modified gravity. In the latter cases, a
space-time volume element must be known before deriving equations of motion
because it is required for the action integral.

In a canonical setting, the Hamiltonian constraint of a given theory is not
uniquely defined but may always be subjected to canonical transformations. In
contrast to classical gravity, there is no distinguished set of phase-space
coordinates because it is not clear from the outset which functions will
determine the space-time geometry. The freedom of performing canonical
transformations must therefore be explored in a systematic derivation of
modified canonical theories. In spherically symmetric models of emergent
modified gravity up to second order in spatial derivatives, it turns out that
one can fix the freedom of canonical transformations by requiring that the
Hamiltonian constraint is strictly periodic in the momentum $K_{\varphi}$,
which then appears with a constant coefficient $\bar{\lambda}$ in
trigonometric functions \cite{HigherCov}. The set of covariant Hamiltonian
constraints of this form is given by
\begin{widetext}
\begin{eqnarray}
    \tilde{H}^{\rm grav}
    &=& - \bar{\lambda}_0 \frac{\sqrt{E^x}}{2} \Bigg[ E^\varphi \Bigg(
    - \bar{\Lambda}_{0}
    + \frac{\bar{\alpha}_0}{E^x}
    + 2 \frac{\sin^2 \left(\bar{\lambda} K_\varphi\right)}{\bar{\lambda}^2}\frac{\partial c_{f}}{\partial E^x}
    + 4 \frac{\sin \left(2 \bar{\lambda} K_\varphi\right)}{2 \bar{\lambda}} \frac{\partial \bar{q}}{\partial E^x}
    \nonumber\\
    &&\qquad
    + \frac{\bar{\alpha}_2}{E^x} \left( c_f \frac{\sin^2 \left(\bar{\lambda} K_\varphi\right)}{\bar{\lambda}^2}
    + 2 \bar{q} \frac{\sin \left(2 \bar{\lambda} K_\varphi\right)}{2 \bar{\lambda}} \right)
    \Bigg)
    + 4 K_x \left(c_f \frac{\sin (2 \bar{\lambda} K_\varphi)}{2 \bar{\lambda}}
    + \bar{q} \cos(2 \bar{\lambda} K_\varphi)\right)
    \nonumber\\
    &&\qquad
    - \frac{((E^x)')^2}{E^\varphi} \left(
    \frac{\bar{\alpha}_2}{4 E^x} \cos^2 \left( \bar{\lambda} K_\varphi \right)
    - \frac{K_x}{E^\varphi} \bar{\lambda}^2 \frac{\sin \left(2 \bar{\lambda} K_\varphi \right)}{2 \bar{\lambda}}
    \right)
    + \left( \frac{(E^x)' (E^\varphi)'}{(E^\varphi)^2}
    - \frac{(E^x)''}{E^\varphi} \right) \cos^2 \left( \bar{\lambda} K_\varphi \right)
    \Bigg]
    \label{eq:Hamiltonian constraint - Anomaly-free - covariant - spherical - vacuum - general - Fully canonically moded - NO lambda}
\end{eqnarray}
such that the structure function equals
\begin{equation}
    \tilde{q}^{x x}
    =
    \left(
    \left( c_{f}
    + \left(\frac{\bar{\lambda} (E^x)'}{2 E^\varphi} \right)^2 \right) \cos^2 \left(\bar{\lambda} K_\varphi\right)
    - 2 \bar{q} \bar{\lambda}^2 \frac{\sin \left(2 \bar{\lambda} K_\varphi\right)}{2 \bar{\lambda}}\right)
    \bar{\lambda}_0^2 \frac{E^x}{(E^\varphi)^2}
    \,.
    \label{eq:Structure function - Anomaly-free - covariant - spherical - vacuum - general - Fully canonically moded - NO lambda}
  \end{equation}
\end{widetext}
The parameter $\bar{\lambda}$ is constant, while
$\bar{\lambda}_0$, $\bar{\Lambda}_0$, $\bar{\alpha}_0$, $\bar{\alpha}_2$,
$c_f$ and $\bar{q}$ are free functions of $E^x$ that define a specific
covariant modification of spherically symmetric general relativity. The
functions $\bar{\Lambda}_0$ and $\bar{\alpha}_0$ can be combined to a dilaton
potential and may also be realized in an action principle. There is no known
action principle for the general set of modifications. Compared with
non-covariant studies of LTB models with modifications from loop quantum
gravity, the last two lines of (\ref{eq:Hamiltonian constraint - Anomaly-free
  - covariant - spherical - vacuum - general - Fully canonically moded - NO
  lambda}) are new and crucial for covariance.
 
This form is useful as a mathematical condition to fix the freedom of gauge
transformations, and also makes it possible to realize the Hamiltonian
constraint as an operator in loop quantum gravity where periodic functions of
$K_{\varphi}$ can be related to basic holonomy operators. However, it is not
always the best version in phenomenological evaluations of the theory because the equations of motion and a relation to physical quantities may simplify in different phase space coordinates \cite{EmergentMubar}. For
this purpose, one may apply a canonical transformation to new phase-space
variables such that $\bar{\lambda}K_{\varphi}$ is replaced by
$\lambda(E^x)K_{\varphi}$ with a scale-dependent function $\lambda(E^x)$. A
canonical transformation then transforms $E^{\varphi}$ to
$(\bar{\lambda}/\lambda(E^x))E^{\varphi}$, while $E^x$ is unchanged and $K_x$
is mapped to a specific linear combination of $K_x$ and $K_{\varphi}$ with
triad-dependent coefficients.
In this form, the nonperiodic modified constraint is
\begin{widetext}
\begin{eqnarray}
    \label{eq:Hamiltonian constraint - modified - non-periodic}
    \!\!&&\!\!\tilde{H}^{\rm grav}
    = - \sqrt{E^x} \frac{\lambda_0}{2} \bigg[ E^\varphi \bigg( -\Lambda_0
    + \frac{\alpha_0}{E^x}
    + 2 \frac{\sin^2 \left(\lambda K_\varphi\right)}{\lambda^2}\frac{\partial c_{f}}{\partial E^x}
    + 4 \frac{\sin \left(2 \lambda K_\varphi\right)}{2 \lambda} \frac{1}{\lambda} \frac{\partial \left(\lambda q\right)}{\partial E^x}
    \\
    \!\!&&
    + \left(\frac{\alpha_2}{E^x} - 2 \frac{\partial \ln \lambda^2}{\partial E^x}\right) \left( c_f \frac{\sin^2 \left(\lambda K_\varphi\right)}{\lambda^2}
    + 2 q \frac{\sin \left(2 \lambda K_\varphi\right)}{2 \lambda} \right)
    + 4 \left(\frac{K_x}{E^\varphi} + \frac{K_\varphi}{2} \frac{\partial \ln \lambda^2}{\partial E^x} \right) \left(c_f \frac{\sin (2 \lambda K_\varphi)}{2 \lambda}
    + q \cos(2 \lambda K_\varphi)\right)
    \bigg)
    \nonumber\\
    \!\!&&
    - \frac{((E^x)')^2}{E^\varphi} \bigg(
    \frac{\alpha_2}{4 E^x} \cos^2 \left( \lambda K_\varphi \right)
    - \left( \frac{K_x}{E^\varphi} + \frac{K_\varphi}{2} \frac{\partial \ln \lambda^2}{\partial E^x} \right) \lambda^2 \frac{\sin \left(2 \lambda K_\varphi \right)}{2 \lambda} \bigg)
    + \left(\frac{(E^x)' (E^\varphi)'}{(E^\varphi)^2}
    - \frac{(E^x)''}{E^\varphi}\right) \cos^2 \left( \lambda K_\varphi \right)
    \bigg]
    \nonumber
\end{eqnarray}
with the associated  structure function 
\begin{eqnarray}
    \tilde{q}^{x x} &=&
    \left(
    \left( c_f
    + \lambda^2 \left( \frac{(E^x)'}{2 E^\varphi} \right)^2
    \right)
    \cos^2 \left( \lambda K_\varphi \right)
    - 2 q \lambda^2 \frac{\sin (2 \lambda K_\varphi)}{2 \lambda}
    \right) \lambda_0^2
    \frac{E^x}{(E^\varphi)^2}
    \,.
    \label{eq:Structure function - modified - non-periodic}
\end{eqnarray}
\end{widetext}
(This expression resembles (\ref{eq:Structure function -
    Anomaly-free - covariant - spherical - vacuum - general - Fully
    canonically moded - NO lambda}), but it is a different phase-space
  function because $\lambda$, unlike $\bar{\lambda}$, in general depends on
  $E^x$. For other canonical transformations, there may be more pronounced
  differences in the expression for the radial metric. Nevertheless, the
  geometrical information contained in the corresponding line element is invariant.)
The remaining modification functions are related by
\begin{eqnarray}
    &&\!\!\lambda_0 = \bar{\lambda}_0 \frac{\lambda}{\bar{\lambda}}
    \quad,\quad
    q = \bar{q} \frac{\bar{\lambda}}{\lambda}
    \quad,\quad
    \Lambda_{0} = \frac{\bar{\lambda}^2}{\lambda^2} \bar{\Lambda}_{0}
    \,,
    \nonumber\\
    &&\!\!
    \alpha_0 = \frac{\bar{\lambda}^2}{\lambda^2} \bar{\alpha}_0
    \quad,\quad
    \alpha_2 = \bar{\alpha}_2 + 4 E^x \frac{\partial \ln \lambda}{\partial E^x}
    \,.
    \label{eq:Redefinitions of lambda}
  \end{eqnarray}

The constraint (\ref{eq:Hamiltonian constraint - Anomaly-free - covariant - spherical - vacuum - general - Fully canonically moded - NO lambda}) has the weak observable
\begin{widetext}
\begin{eqnarray}
    \mathcal{M} &=&
    d_0
    + \frac{d_2}{2} \left(\exp \int {\rm d} E^x \ \frac{\bar{\alpha}_2}{2 E^x}\right)
    \left(
    c_f \frac{\sin^2\left(\bar{\lambda} K_{\varphi}\right)}{\bar{\lambda}^2}
    + 2 \bar{q} \frac{\sin \left(2 \bar{\lambda}  K_{\varphi}\right)}{2 \bar{\lambda}}
    - \cos^2 (\bar{\lambda} K_\varphi) \left(\frac{(E^x)'}{2 E^\varphi}\right)^2
    \right)
    \nonumber\\
    &&
    + \frac{d_2}{4} \int {\rm d} E^x \ \left( \left(
    \bar{\Lambda}_{0}
    + \frac{\bar{\alpha}_0}{E^x}
    \right) \exp \int {\rm d} E^x \ \frac{\bar{\alpha}_2}{2 E^x}\right)
    \label{eq:Gravitational weak observable}
\end{eqnarray}
\end{widetext}
which can be transformed canonically to the observable relevant for
(\ref{eq:Hamiltonian constraint - modified - non-periodic}).  These
expressions can be related to the mass and are therefore useful in the context
of LTB collapse.
A physical system is invariant under canonical transformations and, therefore, both versions, periodic and nonperiodic, predict the same dynamical solutions. In what follows, however, we will use the nonperiodic version (\ref{eq:Hamiltonian constraint - modified - non-periodic}) because it turns out to yield the simpler equations of motion.

Including perfect-fluid degrees of freedom, the general form of the matter
contribution to the Hamiltonian constraint 
is given by \cite{EmergentFluid}
\begin{eqnarray}
    \!\tilde{H}^{\rm matter}_s \!\!\!&=&\!\!\!
    \sqrt{ s P_T^2 + \tilde{q}^{x x} (H_x^{\rm matter})^2}
    - \sqrt{\det{\tilde{q}}} P_q \left(E^x , \frac{P_X^2}{P_T^2}\right)
    \nonumber\\
    \!\!&&
    - 4\pi \sqrt{E^x} E^\varphi \bar{\lambda}_0 P_0 (E^x , P_X^2/P_T^2)
    \label{eq:Hamiltonian constraint - PF - spherical - EMG}
\end{eqnarray}
with the classical contribution (\ref{eq:Diffeomorphism constraint - PF -
  spherical}) to the diffeomorphism constraint. In \cite{EmergentFluid} only
the pressure $P_q$ was introduced, but it is straightforward to check that the
alternative pressure function $P_0$ preserves all the symmetry requirements to
be imposed. A perfect fluid may therefore couple to the emergent spatial
volume element, $\sqrt{\tilde{q}}$, or the classical volume element
$\sqrt{E^x}E^{\varphi}$.  This modified constraint is obtained from requiring
anomaly freedom, normalization of the fluid's velocity co-vector
\begin{equation}
    u_\mu {\rm d} x^\mu = - \left(
      \dot{T} + \frac{P_X}{P_T} \dot{X} \right) {\rm d} t 
    - \left( T' + \frac{P_X}{P_T} X' \right) {\rm d} x \,,
\end{equation}
derived in (\ref{eq:Perfect fluid covector fieldApp}), and the matter covariance condition
\begin{equation}\label{eq:Matter covariance condition}
    \delta_\epsilon u_\mu |_{\rm O.S.} = \mathcal{L}_\xi u_\mu |_{\rm O.S.}
    \,.
\end{equation}

As shown in \cite{SphSymmEff,SphSymmEff2,SphSymmEff3} for constant $\lambda(x)$ and in \cite{EmergentMubar} for general $\lambda(x)$ in the vacuum case, a
non-zero choice for the modification function $\lambda(x)$ implies that the classical
singularity of spherically symmetric solutions is replaced by a transition
between a black hole and a white hole. These two space-time regions are
connected at a maximum-curvature hypersurface defined by $\lambda
K_{\varphi}=\pi/2$ at a local maximum of $\sin(\lambda K_{\varphi})$. While
$K_{\varphi}$ could take greater values, this phase-space function no longer
represents extrinsic curvature in a modified theory. Instead, the space-time
curvature of the resulting emergent metric shows that a maximum is reached at
the given value of $\lambda K_{\varphi}$.

This result has been extended to non-vacuum solutions for dust
\cite{EmergentFluid}, electromagnetic fields \cite{SphSymmEff3,EmergentEM} and
scalar fields \cite{EmergentScalar}. In these cases, however, singularity
resolution is not generic and therefore appears to be unstable under the
inclusion of matter.  In particular, the non-singular vacuum model may be used
as a background for matter in a version of minimal coupling
\cite{SphSymmMinCoup}, but it re-introduces a physical singularity in collapse
models \cite{EmergentScalar}.  (Additional modification functions in matter
terms can be considered that may be preferred for various physical reasons, as
discussed in detail in \cite{EmergentScalar}. They can help to recover the
nonsingular behavior, but their equations of motion differ from minimal
coupling.)  By extension, models of loop quantum gravity are also subject to
the caveat that covariant matter couplings may undo singularity resolution
observed in vacuum models. These results provide strong motivation for a
detailed evaluation of LTB models in emergent modified gravity.

\subsection{LTB conditions}

Because the LTB conditions are phenomenological and not fundamental, since
they merely restrict attention to a convenient set of solutions, there is no
systematic derivation of new conditions in a modified theory. The classical
motivation for the specific condition relies on the resulting line element,
but in emergent modified gravity (or any covariant model of loop quantum
gravity), the complete space-time metric in terms of phase-space variables is
available only after constraints and equations of motion have been analyzed in
order to impose covariance. This feature results in a new ambiguity because
the phase-space form of the LTB condition is no longer equivalent to the
space-time form derived from the line element. In particular, if we maintain
the phase-space condition (\ref{eq:LTB constraint - GR}), the resulting
space-time solutions will not have a simple radial component given by
$(R')^2$, but also contain an additional factor from $\tilde{q}_{xx}$ that is
in general not constant. Alternatively, one may try to derive a condition such
that the new $\tilde{q}_{xx}$ is of the simple form $(R')^2$ (in the marginal
case of $\kappa(x)=0$), which would then require a modified time dependence of
$R$ derived from the resulting equations of motion. However, this procedure
would not only be more complicated, it might also be too restrictive because
it is not guaranteed that such solutions exist for sufficiently general
modification functions.

An additional subtlety is that the phase-space version of the LTB condition
depends on the canonical variables in which the theory is evaluated. For
instance, if we use the classical phase-space condition $L_1$ with a choice of
phase-space variables that implies the non-periodic version of the Hamiltonian
constraint (\ref{eq:Hamiltonian constraint - modified - non-periodic}), the resulting solutions correspond to a periodic theory, with constraint (\ref{eq:Hamiltonian constraint - Anomaly-free - covariant - spherical - vacuum - general - Fully canonically moded - NO lambda}), in which
the transformed LTB condition
\begin{equation}\label{eq:LTB constraint - EMG - periodic}
    \tilde{L}_1 = \frac{\lambda(E^x)}{\bar{\lambda}} E^\varphi - \frac{(E^x)'}{2}
\end{equation}
has been used. These two theories are equivalent, and hence the periodic version of the LTB condition $\tilde{L}_1$ is canonically equivalent to the original $L_1$, given by Eq.~(\ref{eq:LTB constraint - GR}), if one uses, respectively, the periodic and nonperiodic versions of the Hamiltonian constraint.
However, using the periodic
version of the constraint with $\tilde{L}_1$ is not the same as using the periodic version of the constraint with the original $L_1$, as doing so implies one has not connected the two versions of the LTB condition by the corresponding canonical transformation.
We are not aware of any physical condition that could
relate the choice of phase-space variables to a specific phase-space
expression for $L_1$. Again, the specific implementation of LTB solutions
within a covariant modified theory is mainly a matter of convenience.

Here, we maintain the LTB gauge (\ref{eq:LTB gauge}) as well as the
phase-space version (\ref{eq:LTB constraint - GR}) of the LTB condition in the
non-periodic version of the modified Hamiltonian constraint, working with the
original function $L_1$. This choice is appropriate because the non-periodic
constraint and solutions in LTB form are motivated mainly by phenomenological
rather than fundamental reasoning.  We then have to re-evaluate the secondary
condition that $L_1$ be preserved under modified evolution. This condition may
well place restrictions on the modification functions of the new constraint
because there could be covariant modified theories that do not admit solutions
in LTB form.  The resulting equations become very complicated in the modified
theory for non-trivial $\lambda$.  However, the marginally bound case
$\kappa(x)=0$ is tractable and considerable analytic progress is possible.  We
will therefore restrict our attention to this case in the following.

If we use the unmodified LTB condition (\ref{eq:LTB constraint - GR}), the
secondary condition
$\tilde{L}_2 = \dot{L}_1 = \{L_1 , \tilde{H}^{\rm grav}[N] + H_x^{\rm grav}
[N^x]\}$ takes the form
\begin{eqnarray}
    \tilde{L}_2 |_{{\rm O.S.} , N=1,N^x=0, L_1=0} &=& 
    \gamma \left( \alpha_2-1 - 2 E^x \frac{\partial \ln \lambda_0}{\partial E^x}\right)
    \,,
\end{eqnarray}
where we have used the nonperiodic version of the Hamiltonian constraint in
vacuum, given by (\ref{eq:Hamiltonian constraint - modified - non-periodic}),
and defined
\begin{equation}\label{eq:Gamma factor in LTB secondary condition}
    \gamma =
    \frac{\lambda_0 (E^x)'}{2 \sqrt{|E^x|}}
    \left(\left(c_f + \lambda^2\right) \frac{\sin (2 \lambda K_\varphi)}{2 \lambda}
    + 2 q \cos (2 \lambda K_\varphi)\right)
    \,.
\end{equation}
The gamma factor has the non-vanishing classical limit $\gamma\to (E^x)' K_\varphi/(2 \sqrt{|E^x|})$ and therefore the secondary condition implies the equation
\begin{equation}
    \alpha_2-1 - 2 E^x \frac{\partial \ln \lambda_0}{\partial E^x} = 0
    \,.
\end{equation}
Thus, there is indeed a requirement on the modification functions for a
consistent LTB form to exist. This condition can be solved for
\begin{eqnarray}\label{eq:LTB secondary condition on mod functions - EMG - vacuum}
    \alpha_2 = 1 + 2 E^x \frac{\partial \ln \lambda_0}{\partial E^x}
    \,,
\end{eqnarray}
which has the correct classical limit $\alpha_2\to1$ as
$\partial \lambda_0/\partial E^x \to 0$.  If this equation for the
modification functions is satisfied, which would be the case, for instance, if we use the
classical values $\lambda_0=1=\alpha_2$, then the LTB condition is consistent
with the equations of motion of the modified theory in vacuum too. There is no
restriction on the other modification functions $\lambda$, $c_f$ and $q$ that
appear in the emergent metric.

As expected, implementing the LTB condition as a phase-space function does not
produce a metric of the classical LTB form because space-time geometry is
determined by a line element with the emergent metric (\ref{eq:ADM line
  element - emergent - spherical}). With the LTB condition, the emergent line
element (\ref{eq:Structure function - modified - non-periodic}) simplifies
slightly to
\begin{eqnarray}
    \tilde{q}^{x x} &=&
    \bigg(
    \left( c_f
    + \lambda^2
    \right)
    \cos^2 \left( \lambda K_\varphi \right)
    \nonumber\\
    &&\qquad\qquad
    - 2 q \lambda^2 \frac{\sin (2 \lambda K_\varphi)}{2 \lambda}
    \bigg) \lambda_0^2
    \frac{E^x}{(E^\varphi)^2}
    \,.
    \label{eq:Structure function - modified - non-periodic simpl}
\end{eqnarray}
If all modification functions other than $\lambda$ and $\lambda_0$ take their classical
values, the non-periodic version of the mass observable (\ref{eq:Gravitational
  weak observable}) can be used to solve for $\cos^2(\lambda K_{\varphi})$,
resulting in the line element
\begin{equation}\label{eq:LTB metric - EMG - nonperiodic}
    {\rm d} s^2     = - {\rm d} t^2 + \left(1 + \lambda^2 \left(1 - \frac{2
  \mathcal{M}}{R} \right) \right)^{-1} \frac{(R')^2}{\lambda_0^2} {\rm d} x^2
  + R^2 {\rm d} \Omega^2 
    .
\end{equation}

We repeat the procedure described above, but this time we include the perfect-fluid contributions. Taking the LTB gauge (\ref{eq:LTB gauge}) and
keeping an unmodified LTB condition (\ref{eq:LTB constraint - GR}), the
secondary condition
$\tilde{L}_2 = \dot{L}_1 = \{L_1 , \tilde{H}^{\rm grav}[N] + \tilde{H}_s^{\rm
  matter}[N] + H_x^{\rm grav} [N^x] + H_x^{\rm matter} [N^x]\}$ now takes the
form
\begin{eqnarray}
    \tilde{L}_2 |_{{\rm O.S.},N=1,N^x=0,L_1=0}
    &=&
    \alpha H_x^{\rm matter}
    + \beta P_q
    \\
    &&
    + \gamma \left( \alpha_2-1 - 2 E^x (\ln \lambda_0)'\right)
    \,.\nonumber
\end{eqnarray}
where $\gamma$ is again given by the expression (\ref{eq:Gamma factor in LTB secondary condition}), and
\begin{equation}
    \beta = 4 \pi \lambda_0 \gamma  \sqrt{|E^x|} \frac{(E^x)^2}{(E^\varphi)^3} \lambda^4 \left(\tilde{q}_{xx} \right)^{3/2}
    \,,
\end{equation}
while $\alpha$ is a more complicated expression.  The secondary condition now
requires the same vacuum condition (\ref{eq:LTB secondary condition on mod
  functions - EMG - vacuum}), but also a vanishing pressure function $P_q=0$.
The remaining factor then simplifies to
\begin{widetext}
\begin{eqnarray}
    \alpha &=& \frac{(E^x)^{3/2}}{\sqrt{E^x} \lambda_0 (E^\varphi)^2} \Bigg(
    \left( \lambda^2 \lambda_0^3 q
    + \lambda^2 \bar{\lambda}_0 \tilde{q}^{xx} \frac{(E^\varphi)^2}{E^x} \frac{\tan (\lambda K_\varphi)}{\lambda} \right) \frac{H_x^{\rm matter}}{\tilde{H}^{\rm dust}_s}
    \nonumber\\
    &&\qquad
    + \tilde{q}^{xx} \left(\frac{(E^\varphi)^2}{E^x}\right)^{3/2} \frac{\cos (2 \lambda K_\varphi)}{\cos^2(\lambda K_\varphi)}
    - 2 \lambda^2 \lambda_0^2 q \frac{E^\varphi}{\sqrt{E^x}} \frac{\tan (\lambda K_\varphi)}{\lambda}
    \Bigg)
\end{eqnarray}
\end{widetext}
where
\begin{equation}
    \tilde{H}_s^{\rm dust} = \sqrt{ s P_T^2 + \tilde{q}^{x x} (H_x^{\rm matter})^2}
    \,.
\end{equation}
Because $\alpha$ does not vanish in the classical limit, the secondary
condition is again satisfied only if $H_x^{\rm matter}=0$, which means that we
must work in a slicing adapted to the fluid's frame and continue to use $s=1$.

The energy density of the perfect fluid in the fluid's frame is given by
\begin{eqnarray}\label{eq:Energy density - EMG}
    \rho = \frac{P_T}{\sqrt{\det \tilde{q}}}\,.
\end{eqnarray}
In LTB coordinates, we may define
\begin{eqnarray}\label{eq:Def m - EMG}
    \rho = \frac{1}{4\pi R^2} \frac{{\rm d} m}{{\rm d} R} = \frac{1}{4\pi R^2} \frac{m'}{R'}\,.
\end{eqnarray}
with some function $m(x,t)$ we refer to as the mass parameter.
Combining (\ref{eq:Energy density - EMG}) and (\ref{eq:Def m - EMG}), we obtain
\begin{equation}\label{eq:PT m' - EMG}
    P_T = \lambda_0 \sqrt{1 + \lambda^2 \left(1 - \frac{2 \mathcal{M}}{R}
      \right)}\; m'
  \end{equation}
  using $\tilde{q}$ from (\ref{eq:LTB metric - EMG - nonperiodic}).

\section{Modified marginal LTB collapse}
\label{s:Collapse}

We are now ready to apply the general theory of emergent modified gravity in
marginal LTB form to the problem of gravitational collapse.
As a summary, the marginally bound LTB condition
\begin{equation}\label{eq:LTB condition - summary - nonperiodic}
    E^\varphi = \frac{(E^x)'}{2}
  \end{equation}
  in the gauge 
\begin{equation}\label{eq:LTB gauge - summary}
    N=1 , \quad N^x=0
    \,,
\end{equation}
can be consistently implemented in the modified theory
provided the following secondary conditions are applied:
\begin{equation}
    \alpha_2 = 1 + 2 E^x \frac{\partial \ln \lambda_0}{\partial E^x}
    \label{eq:LTB secondary condition on mod functions - EMG - vacuum2}
\end{equation}
for two of the gravitational modification functions, $\alpha_2(E^x)$ and $\lambda_0(E^x)$, as well as
\begin{equation}
    P_q = 0
    \label{eq:Pressure q condition - summary}
\end{equation}
for one of the possible pressure functions, and
\begin{equation}
    P_X = \frac{P_T T'}{X'}
    \label{eq:Fluid frame = slicing - summary}
\end{equation}
from $H_x^{\rm matter}=0$ as in the classical case.

In the following we will take the classical values $c_f=\alpha_0=\alpha_2=1$,
$q=0$, and $\Lambda_0=\Lambda$, and further assume a vanishing cosmological
constant and pressure, $\Lambda=0$ and $P_0=0$, in order to study the collapse
of dust in asymptotically flat space-times.  Using these classical values, the
condition (\ref{eq:LTB secondary condition on mod functions - EMG - vacuum})
is satisfied for any constant $\lambda_0\not=0$, which we will keep in the following
equations.  Furthermore, relabeling $E^x=R^2$ and imposing the LTB condition,
the emergent metric (\ref{eq:LTB metric - EMG - nonperiodic}) takes the form
\begin{equation}\label{eq:ADM line element - spherical - EMG - LTB - simp}
    {\rm d} s^2 = - {\rm d} t^2 + \left(1 + \lambda^2 \left(1 - \frac{2 \mathcal{M}}{R} \right) \right)^{-1} \frac{(R')^2}{\lambda_0^2} {\rm d} x^2 + R^2 {\rm d} \Omega^2
\end{equation}
with $\mathcal{M}$ being the vacuum mass observable (\ref{eq:Gravitational
  weak observable}) now taking the form 
\begin{equation}
    \mathcal{M}
    = \frac{R}{2} \frac{1 + \lambda^2}{\lambda^2} \sin^2 (\lambda K_\varphi)
    \label{eq:Weak observable in simple case - nonperiodic}
\end{equation}
as a phase-space function.

The on-shell condition $H_x=0$ gives
\begin{equation}\label{eq:Radial curvature - LTB}
    K_x = \frac{K_\varphi'}{2}
    \,,
\end{equation}
while $\tilde{H}=0$ implies
\begin{eqnarray}\label{eq:On-shell H=0 - EMG}
    m' &=&
    \mathcal{M}'
    \,,
\end{eqnarray}
where we have substituted $P_T=\lambda_0 m'$.
We can therefore directly integrate (\ref{eq:On-shell H=0 - EMG}), obtaining
\begin{equation}
    m = \mathcal{M}
    + m_0 (t)
    \,,
\end{equation}
where $m_0$ is a free function of $t$ but independent of $x$ that we may set to zero such that
$m=\mathcal{M}$. However, it is the latter that enters the space-time metric
and therefore the actual value of $m_0$ is not relevant for most purposes.
Furthermore, it can be shown that
\begin{eqnarray}
    \dot{\mathcal{M}} = 0
    \,,
\end{eqnarray}
and hence the mass function is time independent in this gauge.

\subsection{Equations of motion}

The relevant equations of motion are
\begin{eqnarray}\label{eq:EoM R - EMG - nonperiodic}
    \dot{R} &=&
    \lambda_0 (1+\lambda^2) \frac{\sin (2 \lambda K_\varphi)}{2 \lambda} 
    \,,
\end{eqnarray}
and
\begin{widetext}
\begin{equation}\label{eq:EoM K_phi - EMG - nonperiodic}
    \left( \lambda_0 (1+\lambda^2) \frac{\sin (2 \lambda K_\varphi)}{2 \lambda} \right)^{\bullet} =
    - \frac{\lambda_0^2 (1+\lambda^2)}{2 R} \frac{\sin^2 (\lambda K_\varphi)}{\lambda^2}
    \left(
    (1+\lambda^2) \left(1- R \frac{\partial \ln \lambda}{\partial R}\right) \cos (2\lambda K_\varphi)
    - (\lambda^2-1) R \frac{\partial \ln \lambda}{\partial R}
    \right)
\end{equation}
\end{widetext}
for the gravitational variables, and
\begin{equation}\label{eq:EoM X - EMG - nonperiodic}
    \dot{X} = 0 \quad,\quad
    \dot{T} = {\rm sgn} (P_T) + \frac{T'}{X'} \dot{X}\quad,\quad
    \dot{P}_T = 0
\end{equation}
for the dust degrees of freedom, to which $P_X$ is strictly related by
(\ref{eq:Fluid frame = slicing - summary}). 

The first two equations can be rewritten in terms of the mass function and $R$:
\begin{eqnarray}\label{eq:EoM R - EMG - nonperiodic - in M and R}
    \!\!\!\!\!\!\!\!
    \dot{R} \!\!&=&\!\!
    s_{\rm sin} \lambda_0 \sqrt{\frac{2 \mathcal{M}}{R} \left(1 + \lambda^2 \left(1 - \frac{2 \mathcal{M}}{R}\right)\right)}
    \\
    \label{eq:EoM K_phi - EMG - nonperiodic - in M and R}
    \!\!\!\!\!\!\!\!
    \ddot{R} \!\!&=&\!\!
    - \frac{\lambda_0^2 \mathcal{M}}{R^2} 
    \left(
    1+\lambda^2 \left(1-\frac{4 \mathcal{M}}{R}\right) 
    - \left( R - 2 \mathcal{M} \right) \frac{\partial \lambda^2}{\partial R}
    \right)
\end{eqnarray}
where $s_{\rm sin}= {\rm sgn} \left( \sin (2 \lambda K_\varphi)\right)$.
Taking a time derivative of (\ref{eq:EoM R - EMG - nonperiodic - in M and R}) simply gives (\ref{eq:EoM K_phi - EMG - nonperiodic - in M and R}), and therefore only the former needs to be solved.

The Friedmann-like equation stemming from (\ref{eq:EoM R - EMG - nonperiodic - in M and R}) is
\begin{equation} \label{Friedmann}
    \left(\frac{\dot{R}}{R}\right)^2 = \lambda_0^2 \left(\frac{2 \mathcal{M}}{R^3} \left(1 + \lambda^2 \left(1 - \frac{2 \mathcal{M}}{R}\right)\right)\right)\,.
\end{equation}
This result can be compared with the restriction to the marginal case of an
equation derived in \cite{LTBShock}, which used a constant
$\Delta$ that, by comparing the equations, would be related to our function
$\lambda$ by
\begin{equation}
    \lambda^2 = -\frac{2\mathcal{M}\Delta/R^3}{1-2\mathcal{M}/R}
  \end{equation}
  provided $\lambda_0=1$.  While the Friedmann-like equations might seem equivalent
  in these two cases, the $\Delta$-version would require a specific
  $\lambda$-function that does not agree with the form of holonomy
  modifications that motivated the original terms in \cite{LTBShock}. The
  reason for this discrepancy can be seen in the fact that this reference did
  not implement covariance conditions.  Furthermore, $\lambda$ is a function
  of only $E^x$ according to the covariance conditions for a constraint
  restricted to at most second-order spatial derivatives, while $\mathcal{M}$
  is a complicated phase space function of $K_\varphi$ and $E^\varphi$ and
  $(E^x)'$. (It is a simple function of $x$ only on a given solution but not
  within a modified theory obtained after implementing the $\Delta$-terms.)
  The choice discussed in \cite{LTBShock} therefore is not related to a valid
  covariant formulation.

\subsection{Reflection-symmetry hypersurface}
\label{sec:Reflection symmetry hypersurface}

As discussed in \cite{EmergentFluid,EmergentMubar}, the full constraints
$\tilde{H}=\tilde{H}^{\rm grav} + \tilde{H}_s^{\rm matter}$ and $H_x =
H_x^{\rm grav} + H_x^{\rm matter}$ are symmetric under the time reversal
operation $K_\varphi \to - K_\varphi$, $K_x \to - K_x$, $T \to - T$, $P_X \to
- P_X$ provided $q=0$, which we assume here.
Furthermore, the system possesses a similar symmetry at the phase-space
hypersurface defined by $K_\varphi = - \pi / (2 \lambda)$ and $K_x=T=P_X=0$:
defining $K_\varphi = - \pi/ (2\lambda) - \delta$, the Hamiltonian is
symmetric under the operation $\delta \to - \delta$, $K_x \to - K_x$,
$T \to - T$, $P_X \to - P_X$.
This transformation is a reflection symmetry in the sense that a given
solution to the equations of motion in the region $\delta < 0$ can also be used to describe a solution in the region $\delta > 0$ by a simple reflection. By itself, this argument does not imply that a complete solution around the maximum-curvature hypersurface is strictly reflection symmetric; it only implies that the future part of the solution is the time reversal of some collapsing solution within the complete solutions space. However, in what follows we demonstrate by a detailed analysis of the specific equations to be solved that every single solution must be reflection symmetric.

Given its similarity to time-reversal symmetry, the solution in the region
$\delta > 0$ will look like a time-reversed solution of the region
$\delta < 0$ but it does describe a different region joined at the
reflection-symmetry hypersurface.
The identification and understanding of this symmetry will be useful in
obtaining the global solution since our task is simplified by requiring only
the dynamical solution on one side of the reflection symmetry hypersurface.

\subsection{Generic formation of a singularity}
\label{s:Generic}

Using (\ref{eq:Weak observable in simple case - nonperiodic}),
we find that at the maximum-curvature hypersurface
given by $K_\varphi=-\pi/(2\lambda)$ in space-time we have
\begin{equation} \label{Rsing}
   R = 2 \mathcal{M} \frac{\lambda^2}{1+\lambda^2}
 \end{equation}
as a function of $x$.
This hypersurface replaces $R=0$ in the classical theory, which is obtained in the limit of $\lambda\to0$.

For a monotonically decreasing $\lambda(R)$\textemdash\,such
  that it has a valid large-$R$ behavior\textemdash\,and any mass distribution
  ${\cal M}(R)$ that increases more slowly than $R(1+\lambda^2)/\lambda^2$,
  equation~(\ref{eq:Weak observable in simple case - nonperiodic}) implies
  that decreasing $R$ implies increasing $\sin^2(\lambda K_{\varphi})$, and
  vice versa. This replaces the classical case where $K_\varphi^2$ increases
  only if ${\cal M}(R)$ decreases more slowly than $R$ for decreasing $R$.
The limiting distribution ${\cal M}\propto R(1+\lambda^2)/\lambda^2$ in general increases with $R$ at a much faster rate than the classical limiting distribution ${\cal M}\propto R$, and, therefore, the assumptions of the modified system encompass a larger class of mass distributions than general relativity. If the mass distribution or $\lambda(R)$ are such that they violate the previous assumptions, then the maximum-curvature hypersurface is not reached and the solution can go all the way to $R=0$ rather than a minimum finite value; in such a case, as we will show below, a singularity develops at $R=0$.
If we extend $K_{\varphi}$ across the value $K_\varphi=-\pi/(2\lambda)$, where the sine squared has a local maximum, we may use this variable as a local time coordinate around the reflection-symmetry hypersurface.
Since we can simply reparametrize time along the trajectories of dust, this coordinate change does not affect the spatial coordinate $x$.
 
Therefore, we can take a spatial derivative of (\ref{Rsing}) at constant $K_{\varphi}$:
 \begin{equation}
   R'= \left( \frac{1 + \lambda^2}{2\lambda^2} - \frac{R}{2}
     \frac{\partial \ln \lambda^2}{\partial R} \right)^{-1} \mathcal{M}'\,.
 \end{equation}
 This expression is non-zero for a non-constant mass profile, such that we avoid shell-crossing singularities.
 Equations~(\ref{eq:EoM R
  - EMG - nonperiodic - in M and R}) and (\ref{eq:EoM K_phi - EMG -
  nonperiodic - in M and R}) imply the time derivatives $\dot{R}=0$ and
\begin{equation}
    \ddot{R} =
    \frac{\lambda_0^2 \mathcal{M}}{R^2}
    \left(
    1+\lambda^2
    - 2 \lambda^2 R \frac{\partial \ln \lambda^2}{\partial R}
    \right)
\end{equation}
which allow us to compute the value of the Ricci scalar of the
space-time metric (\ref{eq:ADM line element - spherical - EMG - LTB - simp}) at
the maximum-curvature hypersurface:
\begin{widetext}
\begin{eqnarray}
    \!\!
    \mathcal{R} \!\!&=&\!\! \Omega\Bigg[\frac{\partial\lambda}{\partial R}\Bigg(- 4\lambda^5\mathcal{M}^3R\left(\lambda^2 - 7\right) + 2 \lambda^3\mathcal{M}^2R^2\left(4\lambda^4 + 9\lambda^2 + 1\right)
     - \lambda\mathcal{M}R^2\left(\lambda^2 + 1\right)\left(3\lambda^4 +
       17\lambda^2 + 6\right)+ 2\lambda R^4\left(\lambda^2 +
       1\right)\Bigg)\nonumber\\
  &&- 2\lambda^4\mathcal{M}R^2\left(R - 8\mathcal{M}\right)\left(R - 2\mathcal{M}\right)\left(\frac{\partial \lambda}{\partial R}\right)^2 - 2\lambda^2\mathcal{M}R^3\left(R - 6\mathcal{M}\right)\left(\frac{\partial\lambda}{\partial R}\right)^2
    \nonumber\\
    &&
    +\left(1+\lambda^2\right)\Bigg(4\lambda^6\mathcal{M}^3 - 19\lambda^4\mathcal{M}^2R\left(\lambda^2 + 1\right) + 2\lambda^2\mathcal{M}R^2\left(4\lambda^4 + 9\lambda^2 + 5\right)
    \nonumber\\
    &&
    +2\lambda^3\mathcal{M}R^2\left(2\mathcal{M} - R\right)\left(2\lambda^2\mathcal{M} - \left(\lambda^2 + 1\right)R\right)\frac{\partial^2 \lambda}{\left(\partial R\right)^2} - \left(\lambda^2 + 1\right)^2R^3\Bigg) + \frac{R^2}{\lambda^2_0}\Bigg] \nonumber
\end{eqnarray}
\end{widetext}
where
\begin{equation}
\Omega = \frac{2\lambda_0^2}{\lambda^2\left(1 + \lambda^2\right)R^4\left(R + \lambda^2\left(R - 2\mathcal{M}\right)\right)}
\end{equation}
This expression is generically divergent at the maximum-curvature hypersurface for
any $\lambda$ because the overall denominator in the last term diverges,
$\left(R + \lambda^2 (R-2 {\cal M})\right)^{-1} \to \infty$ according to (\ref{Rsing}).
It also diverges at $R=0$ if the mass distribution is such that the maximum-curvature hypersurface is not reached.

Following the discussion of Subsection~\ref{sec:Reflection symmetry hypersurface}, the maximum-curvature hypersurface corresponds to the hypersurface of reflection symmetry.
Therefore, the region beyond this hypersurface will be described by a family of reflected collapse solutions, one for each particle trajectory (at least locally, until a shell-crossing singularity is reached).
Since the collapse of $R$ is governed by a single first-order differential equation in time along a dust particle trajectory and the final value of collapse agrees with the initial value of subsequent expansion, a single solution is strictly reflection symmetric with expansion given precisely by the time reversal of collapse along the same trajectory.
The global space-time can then be understood as a collapsing phase that bounces into an expanding phase, both joined at the singular maximum-curvature hypersurface defined by (\ref{Rsing}).
This observation rules out the formation of shock waves if there are no shock profiles present in the initial data of collapse in a region around high curvature. (It is possible that shock waves may form at lower curvature before the region around reflection symmetry is reached. However, such effects would be unrelated to a potential bouncing property of the interior black-hole region.)  While the collapsing phase describes the final stage of a black hole, the expanding phase describes the formation of a white hole. Unlike in vacuum solutions, these two regions are not causally connected but rather appear on two sides of a physical singularity. They may be connected only if one goes beyond an effective description and uses a microscopic theory of quantum space-time. However, possible results of such a transition are beyond the capabilities of current modeling in loop quantum gravity.

For a given space-time solution, it is possible to eliminate the singularity
by choosing a specific modification function
\begin{equation}
  \lambda_0\propto \sqrt{1+\lambda^2(1-2{\cal M}/R)}
\end{equation}
as suggested in \cite{SphSymmLTB2}. (Alternatively, this choice could be
implemented by applying a conformal transformation of the entire emergent
metric by $\omega^2=1+\lambda^2(1-2{\cal M}/R)$ \cite{SphSymmLTB1}, but a
conformal transformation changes the underlying physics if the new solution is
interpreted in the observer's frame; see also \cite{HypDef} and
App.~\ref{a:Conformal} for additional details.) According to (\ref{eq:LTB
  secondary condition on mod functions - EMG - vacuum}), a non-trivial choice
would then be required for $\alpha_2$. While there may be a covariant modified
theory that renders solutions with a given ${\cal M}$ non-singular, solutions
with other mass distributions would remain singular. It is not possible to use
an ${\cal M}$-dependent (and thus solution-dependent) modification function
$\lambda_0$ in a single modified theory. Using the expression (\ref{eq:Weak
  observable in simple case - nonperiodic}), the ${\cal M}$-dependence can be
turned into a specific phase-space dependence on $R$ and $K_{\varphi}$ on the
entire subspace of LTB solutions. However, if the LTB condition is not used,
considering the full modified theory, ${\cal M}$ depends also on $(E^x)'$,
such that the required $\lambda_0$ has a non-polynomial dependence on spatial
derivatives of $E^x$. From the perspective of effective field theory,
derivations within such a model would not be considered reliable because the
remaining terms in the modified Hamiltonian constraint have been restricted to
at most second order in spatial derivatives. It is then not clear how
potential higher spatial derivatives in the other terms of the Hamiltonian
constraint could affect properties of solutions.

We conclude that the covariant collapse of dust with holonomy modifications as
suggested by loop quantum gravity is generically singular for solutions in
marginal LTB form, which is the same conclusion that was reached in
\cite{EmergentFluid} for Oppenheimer-Snyder collapse.

\subsection{Collapse in Gullstrand-Painlev\'{e} coordinates and causal structure}

Even though solutions are generically singular at the maximum-curvature
hypersurface, it is of instructive to compare their causal structure to
classical collapse.  We have two cases of interest: 1) The simple case of
constant holonomy parameter $\lambda=\bar{\lambda}$, and 2) the
scale-dependent holonomy $\lambda(E^x) = \sqrt{\Delta/E^x}$. As already
mentioned, it is possible to use a canonical transformation in order to map
the $K_{\varphi}$-dependence of the Hamiltonian constraint with constant
$\bar{\lambda}$ to an equivalent version with non-constant
$\lambda(E^x)$. However, this transformation, in general, changes the other
modification functions according to (\ref{eq:Redefinitions of lambda}); see
\cite{Lessons} for a detailed discussion. If we make different choices for
$\lambda$ while keeping the other modification functions unchanged, we obtain
physically inequivalent systems.

We will first transform the LTB gauge of space-time solutions with $N^x=0$ to
a Gullstrand--Painlev\'{e} (GP) gauge which is useful for studying collapse across
the point of horizon formation. The relevant equations can be derived for
generic $\lambda$ and will then be specialized and solved in the two main
cases of interest.

To this end, it is convenient to choose $R$ as our new radial coordinate, such that
\begin{eqnarray}
    {\rm d} R &=& R' {\rm d} x + \dot{R} {\rm d} t
    \nonumber\\
    &=& R' {\rm d} x + s_{\rm sin} \lambda_0 \sqrt{\frac{2 \mathcal{M}}{R} \left(1 + \lambda^2 \left(1 - \frac{2 \mathcal{M}}{R}\right)\right)} {\rm d} t
\end{eqnarray}
can be solved for
\begin{equation}
    {\rm d} x = \frac{{\rm d} R}{R'} - s_{\rm sin} \lambda_0 \sqrt{\frac{2 \mathcal{M}}{R} \left(1 + \lambda^2 \left(1 - \frac{2 \mathcal{M}}{R}\right)\right)} \frac{{\rm d} t}{R'}\,.
  \end{equation}
The line element then takes the form
\begin{widetext}
\begin{equation}\label{eq:LTB metric - modified - GP}
    {\rm d} s^2 = - {\rm d} t^2 + R^2 {\rm d} \Omega^2
    + \lambda_0^{-2} \left(1 + \lambda^2 \left(1 - \frac{2 \mathcal{M}}{R} \right) \right)^{-1} \left({\rm d} R - s_{\rm sin} \lambda_0 \sqrt{\frac{2 \mathcal{M}}{R} \left(1 + \lambda^2 \left(1 - \frac{2 \mathcal{M}}{R}\right)\right)} {\rm d} t\right)^2
    \,.
    \nonumber
\end{equation}
\end{widetext}
In these coordinates we have a time-dependent mass distribution, $\mathcal{M}(x)=\mathcal{M}(x (t,R))$.

Gullstrand-Painlev\'{e} coordinates can be derived by using properties of
infalling light rays.
The lightlike condition on the tangent vector implies the velocity
\begin{equation}\label{eq:Null ray velocity - modified - GP}
    \frac{{\rm d} R}{{\rm d} t} \Bigg|_{\rm null}
    =
    \lambda_0 \sqrt{1 + \lambda^2 \left(1 - \frac{2 \mathcal{M}}{R} \right)}
        \left( s_n + s_{\rm sin} \sqrt{\frac{2 \mathcal{M}}{R}} \right) 
    \,,
\end{equation}
where $s_n=\pm 1$ is the sign of the square root with $s_n=1$ being the outgoing
ray and $s_n=-1$ the ingoing ray.   During the collapse stage we have $s_{\rm sin}=-1$
and hence the ingoing ray always has a negative radial velocity, whereas the
outgoing ray has a vanishing velocity at the coordinates solving the equation
\begin{equation}\label{eq:Causal horizon}
    R = 2 \mathcal{M} (t,R)
    \,,
\end{equation}
beyond which it becomes ingoing too.
Both ingoing and outgoing null rays have a vanishing velocity at the maximum-curvature hypersurface.

Canonically, the metric (\ref{eq:LTB metric - modified - GP}) is defined by the GP gauge
\begin{equation}
    N=1 \quad , \quad E^x=R^2 
\end{equation}
in terms of the radial coordinate $R$, together with the condition
\begin{equation}
    E^\varphi=R^2\,,
\end{equation}
which may be seen as an implication of the marginal LTB condition in the GP
gauge.  Assuming the classical values $c_f=\alpha_0=\alpha_2=1$, $q=0$, and
$\Lambda_0=\Lambda$, and further choosing a vanishing cosmological constant and
pressure, $\Lambda=0$ and $P_0=0$, the resulting equations of motion, together
with the on-shell conditions $H_x=0=\tilde{H}$, imply
\begin{eqnarray}
    N^x &=& - \lambda_0 \left(1+\lambda^2\right) \frac{\sin\left(2\lambda K_{\varphi}\right)}{2\lambda}
    \\
    &=& - s_{\rm sin} \lambda_0 \sqrt{1+\lambda^2} \sqrt{\frac{2\mathcal{M}}{R}} \sqrt{1-\frac{2\mathcal{M}}{R}\frac{\lambda^2}{1+\lambda^2}}\,.\nonumber
\end{eqnarray}

In this gauge, the vacuum mass observable takes the form
\begin{equation}
    \mathcal{M} =
    \frac{R}{2} \frac{1+\lambda^2}{\lambda^2} \sin^2\left(\lambda K_{\varphi}\right)
    \label{eq:Weak observable in simple case - nonperiodic - GP}
\end{equation}
and has the time derivative
\begin{eqnarray}
    \dot{\cal M} &=& N^x {\cal M}'
    \\
    &=&
    - s_{\rm sin} \lambda_0 \sqrt{1+\lambda^2} \sqrt{\frac{2\mathcal{M}}{R}} \sqrt{1-\frac{2\mathcal{M}}{R}\frac{\lambda^2}{1+\lambda^2}} {\cal M}'\,.\nonumber
\end{eqnarray}

\subsection{Constant holonomy length}
\label{s:Constant}

Since we use classical values for all modification functions except for
$\lambda$ and assume constant $\lambda_0=1/\sqrt{1+\bar{\lambda}^2}$ (required
for an asymptotically flat vacuum solution), a constant
$\lambda=\bar{\lambda}$ corresponds to a $\mu_0$-type scheme in the old
language of loop quantum cosmology. (See \cite{Lessons} for drawbacks of the
old classification of holonomy schemes and a covariant as well as unified
treatment in the light of emergent modified gravity.)

If we insert a constant $\lambda = \bar{\lambda}$ in equation (\ref{eq:EoM R - EMG - nonperiodic - in M and R}) and use the initial condition $R(t=0,x)=x$ and
$s_{\rm sin}=-1$, we obtain a unique real collapsing solution
\begin{equation}
    R_- = x_{\bar \lambda} \left(\frac{\Xi^{1/3}}{2\bar{\lambda}^{2} x_{\bar\lambda}^{2/3}}
    + \frac{2\bar{\lambda}^2 x_{\bar\lambda}^{2/3}}{\Xi^{1/3}} - 1\right)
\end{equation}
compatible with the classical limit. In this expression, we have the $x$-dependent function
\begin{equation}
    x_{\bar \lambda} = 2 \mathcal{M} \frac{\bar{\lambda}^2}{1+\bar{\lambda}^2}
\end{equation}
and the function
\begin{widetext}
\begin{eqnarray}
    \Xi &=& 4 x_{\bar{\lambda}} \left(\left(a^2+ 2 \bar{\lambda}^6\right) x_{\bar{\lambda}}
    - 3 a\bar{\lambda}^2  \sqrt{1+\bar{\lambda}^2} s t\right)
    + 9 \bar{\lambda}^4 \left(1+\bar{\lambda}^2\right) t^2
    \\
    &&
    + \bigg[16 a^4 x_{\bar{\lambda}}^4
    - 96 a^3 \bar{\lambda}^2 \sqrt{1+\bar{\lambda}^2} st x_{\bar{\lambda}}^3
    + 8 a^2 \bar{\lambda}^4 x_{\bar{\lambda}}^2 \left(27 \left(1+\bar{\lambda}^2\right) t^2+8 \bar{\lambda}^2 x_{\bar{\lambda}}^2\right)
    \nonumber\\
    &&\quad
    - 24 a \bar{\lambda}^6 x_{\bar{\lambda}} \sqrt{1+\bar{\lambda}^2} st \left(9 \left(1+\bar{\lambda}^2\right) t^2+8 \bar{\lambda}^2 x_{\bar{\lambda}}^2\right)
    + 9 \left(1+\bar{\lambda}^2\right) \bar{\lambda}^8 t^2 \left(9 \left(1+\bar{\lambda}^2\right) t^2+16 \bar{\lambda}^2 x_{\bar{\lambda}}^2\right)\Bigg]^{1/2}
    \,.\nonumber
\end{eqnarray}
\end{widetext}
where $s=\pm1$, as well as
\begin{equation}
  a = \frac{A^3-\lambda ^6}{A^{3/2}}
\end{equation}
and
\begin{equation}
    A=\frac{\bar{\lambda}^2}{2 x_{\bar\lambda}} \left(x+x_{\bar\lambda}+\sqrt{(x-x_{\bar\lambda}) (x+3 x_{\bar\lambda})}\right)\,.
\end{equation}

This expression has the correct classical limit,
\begin{equation}\label{eq:Classical R solution}
    \lim_{\bar{\lambda}\to0} R_- = \left(x^{3/2} - s \frac{3}{2} \sqrt{2 \mathcal{M}(x)}\; t\right)^{2/3}
\end{equation}
and is real only for
\begin{equation}
    x > 2 {\cal M}(x) \frac{\bar{\lambda}^2}{1+\bar{\lambda}^2}
\end{equation}
with a lower bound that in a region of constant ${\cal M}$ is always less than the Schwarzschild radius.

Notice that, in this solution, the sign parameter $s=\pm1$ always multiplies $t$, and, therefore, the solution with $s=-1$ can be understood as the
time-reversed version of  $s=+1$.
Hence, we may restrict ourselves to the latter solution.
Furthermore, we note that Eq.~(\ref{eq:EoM R - EMG - nonperiodic - in M and R}) in the case $s_{\rm sin}=-1$, is equivalent to the time-reversed version of the $s_{\rm sin}=+1$ case, differing only by the initial condition.
In sight of this, it is more convenient to solve Eq.~(\ref{eq:EoM R - EMG - nonperiodic - in M and R}) with the initial condition $R(t=0)=x_{\bar\lambda}$, such that we obtain the simpler, and manifestly time-reversal invariant, solution
\begin{widetext}
\begin{eqnarray}
    R &=& x_{\bar\lambda} \Bigg[ 2 \bar{\lambda}^{2/3} x_{\bar\lambda}^{2/3} \left(3 |t| \sqrt{\left(1+\bar{\lambda}^2\right) \left(9 \left(1+\bar{\lambda}^2\right) t^2+16 \bar{\lambda}^2 x_{\bar\lambda}^2\right)}+9 \left(1+\bar{\lambda}^2\right) t^2+8 \bar{\lambda}^2 x_{\bar\lambda}^2\right)^{-1/3}
    \\
    &&
    \qquad
    + \frac{x_{\bar\lambda}^{-2/3}}{2 \bar{\lambda}^{2/3}} \left(3 |t| \sqrt{\left(1+\bar{\lambda}^2\right) \left(9 \left(1+\bar{\lambda}^2\right) t^2+16 \bar{\lambda}^2 x_{\bar\lambda}^2\right)}+9 \left(1+\bar{\lambda}^2\right) t^2+8 \bar{\lambda}^2 x_{\bar\lambda}^2\right)^{1/3}
    - 1\Bigg]\nonumber
\end{eqnarray}
\end{widetext}
with classical limit
\begin{equation}
    \lim_{\bar{\lambda}\to0} R = \left(\frac{3}{2} \sqrt{2 \mathcal{M}(x)}\; t\right)^{2/3}\,.
\end{equation}
In this solution, the singularity and the reflection-symmetry hypersurface correspond to $t=0$.

\subsection{Scale-dependent holonomy function}
\label{s:Scale}

A common scale-dependent choice for the holonomy function is given by
$\lambda=\sqrt{\Delta/E^x}=\sqrt{\Delta}/R$, which provides a covariant
formulation of the traditional $\bar{\mu}$-scheme of loop quantum
cosmology. Asymptotic flatness of the vacuum solution then requires $\lambda_0=1$ \cite{EmergentMubar}.
With this choice, Eq.~(\ref{eq:EoM R - EMG - nonperiodic - in M and R}) is
hard to solve analytically.  However, if we look at the initial stages of the collapse
where $\lambda \ll 1$, we can approximate this equation to
\begin{equation}\label{eq:EoM R - EMG - nonperiodic - approx mu-scheme}
    \mathcal{M}
    \approx \frac{R}{4} \frac{1-s_{\rm cos} \sqrt{1-4 \lambda^2 \dot{R}^2}}{\lambda^2}
    \,,
\end{equation}
where $s_{\rm cos} ={\rm sgn}\left(\cos(2\lambda K_\varphi)\right)$.
Using the scale-dependent holonomy function, we obtain
\begin{equation}
    R (t,x) \approx \left(x^3 \pm 3 \sqrt{2 \mathcal{M}} t \sqrt{x^3-2 \Delta \mathcal{M}} + \frac{9 \mathcal{M} t^2}{2}\right)^{1/3}
    \,,
\end{equation}
where we have imposed the initial condition $R(0)=x$ and used $s_{\rm cos}=+1$,
appropriate for the early stages of the collapse. This solution reduces to the
classical one, (\ref{eq:Classical R solution}), in the limit $\Delta\to0$ and
is valid only for the region $\sqrt{\Delta} / R \ll 1$.  However, for
$\Delta\neq0$ the square root restricts this solution to the region
$x>(2\Delta \mathcal{M})^{1/3}$.  For $\Delta$ of the order of a Planck area,
this would be a very small distance and hence we may simply assume that all
observable physical effects occur for $x\gg (2\Delta \mathcal{M})^{1/3}$ and
expand the term in the square root, simplifying it to
\begin{eqnarray}\label{eq:R - mu-scheme - semiclassical}
    R (t,x) \approx \left(\left(x^{3/2} \pm \frac{3 t}{2}  \sqrt{2 \mathcal{M}} \right)^2
    \mp \frac{3 \Delta t}{2} \left(\frac{2\mathcal{M}}{x}\right)^{3/2}\right)^{1/3}
    .
\end{eqnarray}
This approximation will be sufficient for exploring semiclassical effects at
the early stages of the collapse at large distances.  Taking the collapsing
solution with the two sign choices in (\ref{eq:R - mu-scheme - semiclassical})
given by ($-$, $+$), we find that the modification term slows down the
collapse.

\section{Conclusions}
\label{s:Concl}

We have constructed and analyzed models of marginal LTB collapse with
modifications suggested by loop quantum gravity. Compared with previous
proposals, we have included several key ingredients: We used emergent modified
gravity in order to find space-time covariant theories that include the
desired modifications. Different space-time gauges and slicings can then be
analyzed without changing physical implications. The usual concepts of
space-time curvature, geodesics, and horizons are then unambiguously defined
and allow a meaningful analysis of black-hole properties.

Replacing previous proposals for LTB collapse in loop quantum gravity with
covariant versions does not merely change a few coefficients in equations of
motion. Instead, there are several conceptual differences that require a
careful treatment of defining modified theories and implementing classes of
solutions, for instance of LTB form.  The underlying framework of emergent
modified gravity shows that the crucial invariance condition given by general
covariance can be implemented on the presence of holonomy modifications from
loop quantum gravity only if the classical relationship between the space-time
metric and the original phase-space variables is modified. The specific new
dependence is strictly derived from the theory, specified by a modified
Hamiltonian constraint compatible with covariance conditions, and does not
require additional choices.  However, the form of the new metric changes
either the form of LTB conditions, or the form of the resulting space-time
metric if the classical LTB conditions are used.  Since an LTB reduction is
phenomenological rather than fundamental, there is no preferred choice, except
that some cases may be easier to solve than others.  LTB models in loop
quantum gravity are therefore more ambiguous than their classical versions.

The emergent nature of the space-time metric also means that there are no
distinguished phase-space variables with a direct relationship to the spatial
part of the metric. In this situation, it is important to make sure that the
freedom of performing canonical transformations is properly included in
formulating distinct modified theories. Also here, the framework of emergent
modified gravity has already provided a suitable classification of modified
spherically symmetric theories, up to second order in spatial derivatives of
the phase-space functions. A number of modification functions then
characterizes physically distinct modifications, one of which is directly
relevant for holonomy modifications.

The multitude of different modifications combined with the freedom of choosing
a canonical phase-space representation implies that physical effects are best
analyzed from the perspective of effective field theory. One should then
consider theories on an equal footing that contain all possible terms of a
certain derivative order and obey the required symmetries, given here
by general covariance. In a canonical theory, the derivative order is
determined by spatial derivatives only. General covariance then makes sure
that they are accompanied by suitable time derivatives in the resulting
equations of motion. Within this setting, one may choose specific
modification functions (depending on phase-space degrees of freedom but not
their spatial derivatives) in order to model certain effects such as holonomy
modifications in models of loop quantum gravity.

Our analysis in this paper follows these principles. As a result, marginal LTB collapse in loop quantum gravity implies a physical singularity, rather than just a coordinate singularity from shell-crossing.
Other forms of collapsing systems need not share the same singular fate. In particular, different couplings of scalar matter in spherically symmetric emergent modified gravity can lead to either singular or nonsingular geometries in homogeneous slices \cite{EmergentScalar}. It is, therefore, still an open question whether the collapse of other matter fields, or even of dust beyond the marginal LTB and Oppenheimer-Snyder collapse models or the introduction of pressure, can lead to nonsingular geometries.

\section*{Acknowledgements}

This work was supported in part by NSF grant PHY-2206591.

\begin{appendix}

\section{LTB models in classical general relativity}
\label{a:LTB}

We provide a brief review of the classical formulation of LTB models with an
emphasis of canonical properties as required for emergent modified gravity.

\subsection{Geometric formulation}

In classical general relativity, space-time solutions describing the collapse of
inhomogeneous dust are given by the LTB metric
\begin{equation}\label{eq:LTB metricApp}
    {\rm d} s^2 = - {\rm d} t^2 + \frac{(R')^2}{1+\kappa (x)} {\rm d} x^2 +
    R^2 ({\rm d}\vartheta^2+\sin^2\vartheta{\rm d}\varphi^2)
    \,,
\end{equation}
where the two functions $R(t,x)$ and $\kappa(x)>-1$ are related via Einstein's equation by
\begin{equation}
    \dot{R}^2 = \kappa (x) + \frac{F(x)}{R}
    \,.
\end{equation}
The new function $F$ is implicitly given by the solution to
\begin{equation}
    \frac{F'}{8\pi G R^2 R'} = T_{tt}
    \,.
\end{equation}
If the energy density $T_{tt}$ is written in terms of the Misner--Sharp mass
$m$, it takes the form
\begin{equation}
    T_{tt} = \frac{m'}{4 \pi G R^2 R'}
    \,.
  \end{equation}
  
The last two equations then imply that
\begin{equation}
    m = \frac{1}{2} F(x)
\end{equation}
is a function of the radial coordinate only.
We are therefore left with a single equation of motion,
\begin{equation}
    \dot{R}^2 = \kappa (x) + \frac{2 m(x)}{R}
    \,.
\end{equation}
In the marginally bound case where $\kappa (x) = 0$, this equation can be directly integrated to yield
\begin{equation}
    R (t,x) = \pm \left(  \frac{3}{2} \sqrt{2 m} \left(t-t_i(x)\right)\right)^{2/3}
    \,,
\end{equation}
where $t_i(x)$ is an arbitrary function specifying initial values.
Choosing our initial values by $R|_{t=0}=x$, and the negative sign for collapse, we have
\begin{equation}\label{eq:R solution - classicalApp}
    R (t,x) = \pm \left( x^{3/2} - \frac{3 t}{2} \sqrt{2 m(x)} \right)^{2/3}
    \,.
\end{equation}

The Ricci scalar of (\ref{eq:LTB metricApp}) using (\ref{eq:R solution - classicalApp}) is given by
\begin{eqnarray}\label{eq:Ricci scalar - LTB - marginal - classicalApp}
    \mathcal{R} = \frac{6 m m'}{R^{3/2} \left(\left(R^{3/2}-x^{3/2}\right) m'+3 \sqrt{x} m\right)}
    \,,
\end{eqnarray}
where $m'={\rm d} m/{\rm d} x$.
This spacetime has a geometric singularity at $R=0$ given by the surface
\begin{equation}\label{eq:SingularitySurface}
    t(x)= \frac{2 x^{3/2}}{3\sqrt{2 m(x)}}
    \,.
\end{equation}

There is a further singularity when the second term of the denominator in
(\ref{eq:Ricci scalar - LTB - marginal - classicalApp}) vanishes. Solving for $R$
in this case, we obtain
\begin{eqnarray}
    R(t,x) = \left(x^{3/2} - \frac{3\sqrt{x}m(x)}{m'(x)}\right)^{2/3}.
\end{eqnarray}
Setting this expression equal to (\ref{eq:R solution - classicalApp}) in the
negative case, we derive a second singular surface  given by
\begin{eqnarray}
    t(x) = - \frac{\sqrt{2xm(x)}}{m'(x)}\,.
\end{eqnarray}
Since this function is strictly negative for positive $m'$, it constitutes a
past singularity. For a dust distribution of constant density, $m\propto x^3$
and the singular $t(x)$ is independent of $x$.

\subsection{Canonical formulation}

Emergent modified gravity requires a canonical formulation of the
gravitational field as well as matter ingredients.

\subsubsection{Vacuum LTB models}

Using densitized-triad variables of Schwinger type \cite{Schwinger}, the
spherically symmetric reduction of general relativity as derived in
\cite{SphSymm,SphSymmHam} implies the spherical line element
\begin{equation}\label{eq:ADM line element - spherical - GRApp}
    {\rm d} s^2 = - N^2 {\rm d} t^2 + \frac{(E^\varphi)^2}{E^x} ( {\rm d} x +
    N^x {\rm d} t )^2 + E^x ({\rm d} \vartheta^2+\sin^2\vartheta{\rm d}\varphi^2)
  \end{equation}
  with fields $E^x$ and $E^{\varphi}$. We assume that $E^x>0$, fixing the
  orientation of the spatial triad. The densitized-triad components $E^x$ and
  $E^{\varphi}$ as well as their momenta $K_x$ and $K_{\varphi}$ subject to
  the Hamiltonian constraint
\begin{widetext}
\begin{equation}
    H^{\rm grav} [N] = \int {\rm d} x\ N \bigg[ \frac{((E^x)')^2}{8 \sqrt{|E^x|} E^\varphi}
    - \frac{E^\varphi}{2 \sqrt{|E^x|}}
    - \frac{E^\varphi K_\varphi^2}{2 \sqrt{|E^x|}}
    - 2 K_\varphi \sqrt{|E^x|} K_x
    - \frac{\sqrt{|E^x|} (E^x)' (E^\varphi)'}{2 (E^\varphi)^2} 
    + \frac{\sqrt{|E^x|} (E^x)''}{2 E^\varphi}
    \bigg] =0
    \,,
    \label{eq:Hamiltonian constraint - spherical symmetry - Gauss solvedApp}
\end{equation}
\end{widetext}
and the diffeomorphism constraint
\begin{equation}
    H_x^{\rm grav} [N^x] = \int {\rm d} x\ N^r \left( K_\varphi' E^\varphi - K_x (E^x)' \right)=0
    \, ,
    \label{eq:Diffeomorphism constraint - spherical symmetry- Gauss solvedApp}
\end{equation}
where the primes are radial derivatives. As usual \cite{DiracHamGR,Katz,ADM},
the lapse function $N$ and radial shift vector $N^x$ do not have momenta but
rather can be used to specify the space-time gauge or slicing.

Comparing the two metrics (\ref{eq:LTB metricApp}) and (\ref{eq:ADM line element
  - spherical - GRApp}) we find that the relevant LTB gauge is given by
\begin{equation}\label{eq:LTB gaugeApp}
    N=1 \quad, \quad N^x=0
    \,.
\end{equation}
Moreover, the phase space is further restricted by the LTB condition
\begin{equation}\label{eq:LTB conditionApp}
    E^\varphi = \frac{(E^x)'}{2 \sqrt{1+\kappa (x)}}
    \,.
\end{equation}
We also identify $R^2=E^x$ without imposing a further constraint.

For a given $\kappa(x)$, we can directly enforce the LTB constraint
\begin{equation}\label{eq:LTB constraint - GRApp}
    L_1 \equiv E^\varphi - \frac{(E^x)'}{2 \sqrt{1+\kappa (x)}}
    \,,
\end{equation}
set to vanish dynamically.  For a consistent implementation, this condition
must be preserved under time evolution \cite{LTB,LTBII}, that is,
$L_2\equiv \dot{L_1}=\{L,H[N]+H_x[N^x]\}=0$, thereby obtaining a secondary
condition. (Recall that $\kappa$ is a function of the radial
coordinate only and hence $\dot{\kappa}=0$ so that the brackets do not act on
it.)

We compute
\begin{widetext}
\begin{eqnarray}
    L_2 &=& \dot{E}^\varphi  - \frac{1}{2 \sqrt{1+\kappa (x)}} (\dot{E}^x)'
    \nonumber\\
    &=& - \frac{\delta H[N]}{\delta K_\varphi} + \frac{1}{2\sqrt{1+\kappa
        (x)}}  \left(\frac{\delta H[N]}{\delta K_x}\right)' 
    + \left( N^x E^\varphi - \frac{1}{2 \sqrt{1+\kappa (x)}} N^x (E^x)'\right)'
    \nonumber\\
    &=& - \frac{\delta H[1]}{\delta K_\varphi} + \frac{1}{2 \sqrt{1+\kappa
        (x)}} \left(\frac{\delta H[1]}{\delta K_x}\right)' 
    \nonumber\\
    &=& - \frac{\partial H[1]}{\partial K_\varphi}
    + \left(\frac{\partial H[1]}{\partial K_\varphi'}\right)'
    - \left(\frac{\partial H[1]}{\partial K_\varphi''}\right)''
    + \frac{1}{2 \sqrt{1+\kappa (x)}} \left( \left(\frac{\partial H[1]}{\partial K_x}\right)'
    - \left(\frac{\partial H[1]}{\partial K_x'}\right)''
    + \left(\frac{\partial H[1]}{\partial K_x''}\right)''' \right)
    \,.
\end{eqnarray}
\end{widetext}
Using (\ref{eq:Hamiltonian constraint - spherical symmetry - Gauss solvedApp})
this expression equals
\begin{eqnarray}\label{eq:LTB secondary condition explicitApp}
    L_2 \!\!&=&\!\!
    \frac{E^\varphi K_\varphi}{\sqrt{|E^x|}}
    + 2 \sqrt{|E^x|}  K_x
    - \frac{\left( K_\varphi \sqrt{|E^x|} \right)'}{\sqrt{1+\kappa (x)}}
    \\
    \!\!&=&\!\!
    \frac{K_\varphi}{\sqrt{|E^x|}} \left(E^\varphi - \frac{(E^x)'}{2 \sqrt{1+\kappa (x)}}\right)
    \nonumber\\
    \!\!&&\!\!
    - \frac{\sqrt{|E^x|}}{E^\varphi} \left( \frac{E^\varphi K_\varphi'}{\sqrt{1+\kappa (x)}} - 2 K_x E^\varphi \right)
    \nonumber\\
    \!\!&=&\!\!
    \left( \frac{K_\varphi}{\sqrt{|E^x|}}
    + \frac{\sqrt{|E^x|}}{E^\varphi} 2 K_x \right) L_1
    - \frac{\sqrt{|E^x|}}{E^\varphi} \frac{H_x^{\rm grav}}{\sqrt{1+\kappa (x)}}
    \,.\nonumber
\end{eqnarray}
We therefore find that in the given gauge the LTB condition $L_1=0$ is consistent
with the on-shell canonical dynamics, where $H_x^{\rm grav}=0$.

\subsubsection{Perfect-fluid LTB}

The perfect fluid is a model of a collection of point particles in the space-time
manifold which may locally be described by coordinate functions $T(t,x)$ and
$X(t,x)$. We interpret the time dependence $(T(t,x),X(t,x))$ at fixed $x$ as
the event to which the particle at a position labeled by $x$ has moved from
the initial event $(T(0,x),X(0,x))$. The two functions $T$ and $X$ can
therefore be used as basic fields of a dynamical formulation of perfect
fluids. For a given action or set of canonical constraints, the dynamics will
be valid in a region in which trajectories of individual particles do not
cross. Boundaries of this region for a given theory are determined by
shell-crossing singularities.

For a derivation of the canonical dynamics, we combine spatial fields $T=T(0,x)$ and
$X(0,x)$ with canonical momenta $P_T$ and $P_X$ with units of energy and momentum,
respectively. Both momenta are scalar densities of weight one, which in one
radial dimension transform like components of a co-vector. We have canonical
Poisson brackets
\begin{equation}
    \{ T (x) , P_T (y) \}
    = \{ X (x) , P_X (y) \}
    = \delta (x-y)
    \,.
\end{equation}
Since the two fields in spherical symmetry are spatial scalars, we immediately
obtain the fluid contribution
\begin{equation}
  H_x=P_TT'+P_XX'
\end{equation}
to the diffeomorphism constraint.

The dual role of $(T,X)$ as canonical configuration variables and spacetime coordinates of the fluid can be used to
recognize $H_x$ as the spatial component of the energy-momentum co-vector
$(P_T,P_X)$ transformed to a frame with the spatial coordinate $x$ that
appears in the derivatives. The corresponding time
component in this frame is given by $P_T\dot{T}+P_X\dot{X}$. In general a
general frame $(t,x)$, we therefore have the co-vector field
\begin{equation} \label{PPApp}
P_T{\rm d}T+P_X{\rm d}X= ( P_T\dot{T}+P_X\dot{X}){\rm d}t+ (P_TT'+P_XX'){\rm d}x\,.
\end{equation}
The left-hand side is applies in a co-moving frame where $\dot{T}=1=X'$ and
$T'=0=\dot{X}$. 

In special relativity, the momentum co-vector $(P_T,P_X)$ and the coordinate
differential $({\rm d}T,{\rm d}X)$ are both expected to be proportional to the
relativistic velocity $u^{\mu}$ or its dual. The sum in (\ref{PPApp}) then takes
the form of a scalar product
\begin{equation}
  P_T{\rm d}T+P_X{\rm d}X=p_{\mu}{\rm d}x^{\mu} \propto u_{\mu}\frac{{\rm
      d}x^{\mu}}{{\rm d}\tau}=g_{\mu\nu}u^{\mu}u^{\nu}=-s
\end{equation}
where $s=1$ for massive particles and $s=0$ in the massless case. The
proportionality factor in special relativity is given by the mass, but here we
need a scalar of density weight one. The only available choice that can in
general be assumed to be non-vanishing is given by $P_T$. We therefore define
the velocity co-vector
\begin{eqnarray}\label{eq:Perfect fluid covector fieldApp}
    u_\mu {\rm d} x^\mu &=&
    -\frac{1}{P_T}(P_T{\rm d}T+P_X{\rm d}X)
    \\
    &=& - \left(
      \dot{T} + \frac{P_X}{P_T} \dot{X} \right) {\rm d} t 
    - \left( T' + \frac{P_X}{P_T} X' \right) {\rm d} x \,.\nonumber
\end{eqnarray}
The fluid's velocity (\ref{eq:Perfect fluid covector fieldApp}) is normalized
\begin{equation}
    g^{\mu\nu} u_\mu u_\nu = - s\,,
\end{equation}
with $s=1$ or $s=0$ for a timelike or null fluid, respectively. Therefore,
\begin{equation} \label{normApp}
  ||(P_T,P_X)||^2=-sP_T^2\,.
\end{equation}
The relation
between the fluid's velocity and the configuration variables $T$ and $X$ imply
that the latter are capable of locally coordinatizing the manifold with the
simplest example given by $T'=0,\dot{T}=1$ and $X'=1, \dot{X}=0$ with
$P_X/P_T$ describing the velocity of the timelike fluid's particles.

The Hamiltonian-constraint contribution from the perfect fluid in its general
covariant form is given by \cite{EmergentFluid,brown1993action}
\begin{equation}
    H^{\rm matter}_s =
    \sqrt{ s P_T^2 + \frac{E^x}{(E^\varphi)^2} (H_x^{\rm matter})^2}
    - \sqrt{\det{q}} P \left(\frac{P_X^2}{P_T^2}\right)
    \label{eq:Hamiltonian constraint - PF - spherical - classicalApp}
\end{equation}
with a free pressure function $P$ depending on the velocity,
and the diffeomorphism-constraint contribution by
\begin{equation}
    H_x^{\rm matter} =
    P_T T'
    + P_X X'
    \label{eq:Diffeomorphism constraint - PF - sphericalApp}
  \end{equation}
  as already defined.
The coefficient $s=1$ for a timelike fluid and $s=0$ for a null fluid appears
in the general contribution to the Hamiltonian constraint. This square-root
contribution represents the relativistic energy as the time component of the
vector $(H_s^{\rm matter},H_x^{\rm matter})$ with norm squared $-sP_T^2$.
Including the gravitational field, the complete Hamiltonian constraint is then given
$H = H^{\rm grav} + H^{\rm matter}_s$, and the total diffeomorphism constraint
by $H_x = H^{\rm grav}_x + H^{\rm matter}_x$.  The full Hamiltonian generating
the equations of motion is
$H_T[N,N^x]=H^{\rm grav}[N] + H^{\rm matter}_s[N] + H^{\rm grav}_x[N^x] +
H^{\rm matter}_x[N^x]$.

Because the full diffeomorphism constraint is now given by
$H_x=H_x^{\rm grav} + H_x^{\rm matter}$, imposing the LTB condition, the
secondary condition (\ref{eq:LTB secondary condition explicitApp}) receives the
matter contribution
\begin{equation}
    L_2 |_{{\rm O.S.}, L_1=0} = \frac{\sqrt{E^x}}{E^\varphi} \frac{H^{\rm matter}_x}{\sqrt{1+\kappa(x)}}
    \,.
\end{equation}
Therefore, the secondary condition is satisfied only if $H_x^{\rm matter}=0$,
which means we must work in a slicing adapted to the fluid's frame, and
therefore have
\begin{equation}\label{eq:Fluid frame = slicingApp}
    P_X = - \frac{P_T T'}{X'}
    \,,
\end{equation}
and $s=1$, since this is only possible for a timelike fluid. 

The energy density of the perfect fluid in the fluid's frame is given by
\begin{eqnarray}\label{eq:Energy densityApp}
    \rho = \frac{P_T}{\sqrt{\det q}}\,.
\end{eqnarray}
In the LTB coordinates used in (\ref{eq:LTB metricApp}), we may define
\begin{eqnarray}\label{eq:Def mApp}
    \rho = \frac{1}{4\pi R^2} \frac{{\rm d} m}{{\rm d} R} = \frac{1}{4\pi R^2} \frac{m'}{R'}\,.
\end{eqnarray}
with some function $m(x,t)$ we refer to as the mass parameter.
Combining (\ref{eq:Energy densityApp}) and (\ref{eq:Def mApp}), we obtain
\begin{equation}\label{eq:PT m'App}
    P_T = \frac{m'}{\sqrt{1+\kappa(x)}}\,.
\end{equation}

\subsection{LTB collapse}

The on-shell condition $H=0$ together with
(\ref{eq:LTB condition}) and $R^2=E^x$ implies
\begin{equation}\label{eq:H=0 - classical}
    |P_T| =
    \frac{1}{2 \sqrt{1+\kappa}} \left( \left( 8 \pi P - \Lambda \right) \frac{(R^3)'}{3}
    - (\kappa R)'
    + (R K_\varphi^2)' \right)
\end{equation}
where $P$ is the function introduced in (\ref{eq:Hamiltonian constraint - PF - spherical - classical}) and $\Lambda$ is the cosmological constant. 
The relevant equations of motion are
\begin{equation}\label{eq:EoM R - classical}
    \dot{R} = K_\varphi
\end{equation}
and
\begin{equation}\label{eq:EoM K_phi - classical}
    \dot{K}_\varphi =
    \left(\frac{\Lambda}{2} - 4 \pi P \right) R
    + \frac{\kappa}{2 R}
    - \frac{K_\varphi^2}{2 R}
\end{equation}
for the unrestricted gravitational variables, as well as
\begin{equation}\label{eq:EoM X - classical}
    \dot{X} = \frac{8 \pi R^2}{\sqrt{1+\kappa}} R' \frac{T'}{X'} \frac{1}{P_T} \frac{\partial P}{\partial p}
  \end{equation}
  and
  \begin{equation}    \label{eq:EoM T - classical}
    \dot{T} = {\rm sgn} (P_T) + \frac{T'}{X'} \dot{X}
  \end{equation}
  for the dust coordinates,
  where $p = P_X^2 / P_T^2$, and 
\begin{equation}    \label{eq:EoM PT - classical}
    \dot{P}_T=0
  \end{equation}
  for the free dust momentum after using (\ref{eq:Fluid frame = slicingApp}).

  Using (\ref{eq:PT m'App}) and (\ref{eq:EoM R - classical}), the equations
  (\ref{eq:H=0 - classical}) and (\ref{eq:EoM K_phi - classical}) can be
  rewritten as
\begin{equation}\label{eq:H=0 - classical - in R}
    m' =
    \frac{1}{2} \left( \left( 8 \pi P - \Lambda \right) \frac{(R^3)'}{3}
    + \left(R \dot{R}^2 - R \kappa \right)' \right)
    \,,
\end{equation}
\begin{equation}\label{eq:R - classical - simplified}
    \ddot{R} =
    \left(\frac{\Lambda}{2} - 4 \pi P \right) R
    - \frac{\dot{R}^2-\kappa}{2 R}
    \,.
\end{equation}
Equations (\ref{eq:PT m'App}) and (\ref{eq:EoM PT - classical}) also imply that
$m(x)$ is independent of time.

For dust, $P=0$, equation (\ref{eq:H=0 - classical - in R}) can be integrated to yield
\begin{equation}\label{eq: LTB mass function pde solution}
    m (x) = \frac{R (\dot{R}^2-\kappa)}{2} - \frac{\Lambda}{6} R^3
    + m_0
    \,,
\end{equation}
where $m_0$ is the constant of integration. We may set this constant equal to zero by imposing $m |_{R=0}=0$. Further setting $\Lambda=0$, the equation can be exactly
solved for $R(t)$ in the marginally bound case $\kappa=0$,
\begin{equation}\label{eq:R solution - classical - canonical} 
    R = \left(x^{3/2} \pm \frac{3 t}{2} \sqrt{2 m (x)-m_0}\right)^{2/3}
    \,,
\end{equation}
where we have taken the initial condition $R(t=0,x)=x$ to fix the undetermined
functions of $x$ resulting from solving the equation.
Direct substitution shows that this function solves (\ref{eq:R - classical -
  simplified}) too.  The solution (\ref{eq:R solution - classicalApp}) is
precisely the geometric result with the negative sign chosen for the collapsing
case.

The LTB metric (\ref{eq:LTB metric}) in this marginal case becomes
\begin{equation}\label{eq:LTB metric - marginal}
    {\rm d} s^2 = - {\rm d} t^2 + \frac{x}{R} \left(1 - \frac{t}{2} \sqrt{\frac{2 m}{x}} (\ln m)'\right)^2 {\rm d} x^2 + R^2 {\rm d} \Omega^2
    \,.
\end{equation}
This metric has a coordinate singularity at the time
\begin{eqnarray}\label{eq:LTB coordinate singularity - classical}
    t(x) = 2 \sqrt{\frac{x}{2 m(x)}} \frac{m(x)}{m'(x)}
    \,.
\end{eqnarray}

\subsubsection{Collapse in Gullstrand-Painlev\'e coordinates}

If we use $R$ in (\ref{eq:R solution - classicalApp}) as a new radial coordinate we obtain
\begin{eqnarray}\label{eq:R coordinate transformation - classical}
    {\rm d} x &=& \left( 1 - \sqrt{\frac{2 m}{x}} \frac{m'}{m} \frac{t}{2} \right) \sqrt{\frac{R}{x}} \left( {\rm d} R + \sqrt{\frac{2 m}{R}} {\rm d} t \right)
    \nonumber\\
    &=:& \mathcal{W} \left( {\rm d} R + \sqrt{\frac{2 m}{R}} {\rm d} t \right)
    \,,
\end{eqnarray}
where $m' = {\rm d} m / {\rm d} x$.
Substitution into the LTB metric (\ref{eq:LTB metric}) gives
\begin{equation}\label{eq:LTB metric in GP form - classical}
    {\rm d} s^2 = - {\rm d} t^2 + \left( {\rm d} R + \sqrt{\frac{2 m}{R}} {\rm d} t \right)^2 + R^2 {\rm d} \Omega^2
    \,,
\end{equation}
which is the Gullstrand-Painlev\'e metric with non-constant $m (t,R)$.
This form of the metric is useful to cross the event horizon.
Taking an infalling/outgoing light ray, its velocity in the GP coordinates is given by
\begin{eqnarray}\label{eq:GP light ray velocity - classical}
    \frac{{\rm d} R}{{\rm d} t} = \pm 1 - \sqrt{\frac{2 m}{R}}
    \,.
\end{eqnarray}
The horizon surface is given the vanishing of the outgoing rays's velocity, beyond which it becomes ingoing:
\begin{eqnarray}\label{eq:GP horizon - classical}
    2 m (t,R) = R
    \,.
\end{eqnarray}

\subsubsection{Uniform distribution}

Consider a uniform mass distribution originally contained in a radius $r$ in
LTB coordinates, such that $m (x) = M x^3 / r^3 = \rho_0 x^3$, with total mass
$M<r/2$.  In LTB coordinates, the metric (\ref{eq:LTB metric - marginal}) with
this distribution becomes
\begin{equation}\label{eq:LTB metric - marginal - uniform}
    {\rm d} s^2 = - {\rm d} t^2 + \left(1 - \frac{3 \sqrt{\rho_0}}{\sqrt{2}} t \right)^{4/3} {\rm d} x^2 + R^2 {\rm d} \Omega^2
    \,,
\end{equation}
and the coordinate singularity (\ref{eq:LTB coordinate singularity - classical}) for this distribution occurs at the time
\begin{eqnarray}\label{eq:LTB coordinate singularity - classical - uniform}
    t = \frac{2}{3} \sqrt{\frac{x^3}{2 M}}= \frac{\sqrt{2}}{3 \sqrt{\rho_0}} 
    \,.
\end{eqnarray}
The Ricci scalar (\ref{eq:Ricci scalar - LTB - marginal - classicalApp}) becomes
\begin{eqnarray}\label{eq:Ricci scalar - LTB - marginal - classical - uniform}
    \mathcal{R} = \frac{6 M}{x^3} \frac{x^3}{R^3}
    = \frac{6 M}{x^3} \left(1 - \frac{3 \sqrt{\rho_0}}{\sqrt{2}} t\right)^{-2}
    \,.
\end{eqnarray}
The coordinate singularity agrees with the geometric singularity.

With this distribution we obtain
\begin{eqnarray}
    R (t,x) &=& x \left(1 - \frac{3 \sqrt{\rho_0}}{\sqrt{2}} t\right)^{2/3}
\end{eqnarray}
and the LTB metric (\ref{eq:LTB metric}) takes the form of a flat FLRW spacetime
\begin{equation}\label{eq:FLRW metric - classical - marginal}
    {\rm d} s^2 = - {\rm d} t^2 + a(t)^2 \left( {\rm d} x^2 + x^2 {\rm d} \Omega^2 \right)
\end{equation}
with a matter-dominated scale factor
\begin{equation}\label{eq:FLRW scale factor - classical}
    a(t) = \left(1 - \frac{3 \sqrt{\rho_0}}{\sqrt{2}} t \right)^{2/3}
    \,.
\end{equation}

We can then use (\ref{eq:R solution - classicalApp}) to get
\begin{equation}
    {\rm d} R = \left(1 - \frac{3 \sqrt{\rho_0}}{\sqrt{2}} t \right)^{2/3} {\rm d} x
    - R \sqrt{2 \rho_0} \left(1 - \frac{3 \sqrt{\rho_0}}{\sqrt{2}} t\right)^{-1} {\rm d} t
\end{equation}
and rewrite the metric in Gullstrand--Painlev\'{e} form
(\ref{eq:LTB metric in GP form - classical}):
\begin{equation}
    {\rm d} s^2 = - {\rm d} t^2 + \left( {\rm d} R + \frac{2 R}{3} \left( \frac{\sqrt{2}}{3 \sqrt{\rho_0}} - t\right)^{-1} {\rm d} t \right)^2 + R^2 {\rm d} \Omega^2
    \,.
\end{equation}
Further redefining the time coordinate
\begin{eqnarray}
    \tau = t - \frac{\sqrt{2}}{3 \sqrt{\rho_0}}
\end{eqnarray}
we obtain
\begin{equation}\label{eq:Oppenheimer-Snyder collapse metric in GP form}
    {\rm d} s^2 = - {\rm d} \tau^2 + \left( {\rm d} R + \frac{2 R}{3 \tau} {\rm d} \tau \right)^2 + R^2 {\rm d} \Omega^2
    \,,
\end{equation}
which is the Oppenheimer-Snyder model in Gullstrand--Painlev\'{e} coordinates
for a star collapsing from rest at spatial infinity.

The velocity of an infalling/outgoing light ray (\ref{eq:GP light ray velocity - classical}) is given by
\begin{eqnarray}
    \frac{{\rm d} R}{{\rm d} t} = \pm 1 - \sqrt{\frac{2 M}{r}} \frac{R}{r} \left(1 - \frac{3 t}{2 r} \sqrt{\frac{2 M}{r}}\right)^{-1}
    \,.
\end{eqnarray}
The horizon surface (\ref{eq:GP horizon - classical}) is given by
\begin{eqnarray}\label{eq:GP horizon - classical - Uniform}
    t = \frac{\sqrt{2}}{3 \sqrt{\rho_0}} \left( 1 - \sqrt{\frac{2}{\rho_0}} R\right)
\end{eqnarray}
which differs from the coordinate singularity (\ref{eq:LTB coordinate singularity - classical - uniform}).

The geometric singularity (\ref{eq:SingularitySurface}) is
\begin{equation}\label{eq:Singularity surface - classical - uniform}
    t(x)= \frac{\sqrt{2}}{3\sqrt{\rho_0}}
    \,,
\end{equation}
which intersects the horizon at $R=0$ and can be seen directly from Eq.~(\ref{eq:Ricci scalar - LTB - marginal - classical - uniform}).

\section{Conformal transformations of the emergent metric}
\label{a:Conformal}

Conformal transformations of space-time metrics are a useful tool for various aspects of space-time analysis. However, in such cases the purpose of conformal transformations is to simplify the analysis of the causal structure, not to change physical properties of the spacetime geometry given by an explicit solution.
Their application in recent studies of LTB collapse
\cite{SphSymmLTB1,SphSymmLTB2} in order to remove physical singularities is
conceptually different and requires further discussions, which we present in
this appendix. We focus on the two-dimensional case relevant for the
spherically symmetric reduced theory in these applications.

The study \cite{SphSymmLTB1} defined a conformally modified
two-dimensional emergent metric given by
$\tilde{\mathfrak{g}}_{\mu\nu} = \omega^2 \tilde{g}_{\mu\nu}$, explicitly
\begin{equation}
    \tilde{\mathfrak{g}}_{\mu\nu} {\rm d} x^\mu {\rm d} x^\nu = \omega^2 \left( - N^2 {\rm d} t^2 + \tilde{q}_{xx} ({\rm d} x + N^x {\rm d} t)^2 \right)
    \,.
    \label{eq:ADM line element - emergent - spherical - conformal modification}
\end{equation}
If the covariance condition for the emergent metric (\ref{eq:Covariance condition - emergent}) is satisfied, then the covariance condition for the conformally modified emergent metric
\begin{eqnarray}
    \delta_\epsilon \tilde{\mathfrak{g}}_{\mu \nu} \big|_{{\rm O.S.}} &=&
    \mathcal{L}_{\xi} \tilde{\mathfrak{g}}_{\mu \nu} \big|_{{\rm O.S.}}
    \,,
    \label{eq:Covariance condition - conformal modification}
\end{eqnarray}
requires $\omega$ be a space-time scalar, meaning its gauge transformation
must give the Lie derivative
$\mathcal{L}_\xi \omega = \xi^\mu \partial_\mu \omega$ on-shell.  Explicitly, the
condition becomes
\begin{eqnarray}\label{eq:Conformal cov cond}
    \delta_\epsilon \omega = \xi^t \dot{\omega} + \xi^x \omega'
    \,.
\end{eqnarray}
If this is the case, then (\ref{eq:ADM line element - emergent - spherical -
  conformal modification}) is obviously a covariant tensor.  This is a feature
not exclusive of emergent modified gravity.  In general relativity, a
covariant tensor of any rank multiplied by a scalar is still covariant and of
the same rank.  The question is whether (\ref{eq:ADM line element - emergent -
  spherical - conformal modification}) can be used as a space-time metric
replacing the emergent one, since the space-time metric has several properties
that single it out among other tensors, including off-shell properties.

As shown in \cite{HypDef}, such a conformal transformation is possible and in
accordance with the principles of emergent modified gravity (anomaly-freedom
and covariance) only if the Hamiltonian constraint is rescaled to $\bar{H}$
given by
\begin{equation}\label{eq:Conformal constraint}
    \bar{H} = \omega^{-1} \tilde{H}
\end{equation}
and the conformal factor obeys the off-shell gauge transformation
\begin{equation}
    \{\omega , \tilde{H}[\epsilon^0]+H_x[\epsilon^x]\} = \Omega_0 \epsilon^0 + \omega' \epsilon^x
\end{equation}
for some phase space function $\Omega_0$, with no derivatives of $\epsilon^0$
or $\epsilon^x$, and only if $\omega$ commutes with itself.  In this
particular system, if $\omega$ is a function of only the vacuum gravitational
observable $\mathcal{M}$ and $E^x$, then these properties are satisfied.  The
algebra of $\bar{H}$ with the diffeomorphism constraint then takes the
hypersurface-deformation form
\begin{eqnarray}
    \!\!\!\!\{ H_x [\bar{N}^x] , H_x[\bar{M}^x] \} \!\!&=&\!\! - H_x [\bar{M}^x (\bar{N}^x)'-\bar{N}^x (\bar{M}^x)']
    \label{H_x,H_x bracket - conformal}
    \\
    \!\!\!\!\{ \bar{H} [\bar{N}] , H_x [\bar{M}^x] \} \!\!&=&\!\! - \bar{H}[\bar{M}^x \bar{N}'] 
    \label{eq:H,H_x bracket - emergent - conformal}
    \\
    \!\!\!\!\{ \bar{H} [\bar{N}] , \bar{H}[\bar{M}] \} \!\!&=&\!\! - H_x \left[ \bar{q}^{x x} \left( \bar{M} \bar{N}'-\bar{N} \bar{M}' \right)\right]
    \label{eq:H,H bracket - emergent - conformal}
\end{eqnarray}
with structure function
\begin{equation}
    \bar{q}^{x x} = \omega^{-2} \tilde{q}^{x x}\,.
\end{equation}
The emergent line-element compatible with the new algebra is given by
\begin{eqnarray}\label{eq:Conformal metric in conformal frame}
    \bar{g}_{\mu\nu} &=& - \bar{N}^2 {\rm d} t^2 + \bar{q}_{xx} \left({\rm d} x + \bar{N}^x {\rm d} t\right)^2
    \nonumber\\
    &=& - \bar{N}^2 {\rm d} t^2 + \omega^2 \tilde{q}_{xx} \left({\rm d} x + \bar{N}^x {\rm d} t\right)^2
    \,.
\end{eqnarray}

The conformally transformed geometry is therefore anomaly-free and covariant,
in accordance with the principles of emergent modified gravity.  However,
there is an important conceptual distinction between the use of the tensors
(\ref{eq:ADM line element - emergent - spherical - conformal modification})
and (\ref{eq:Conformal metric in conformal frame}) as follows.

In the former
case, the spacetime geometry is given by the emergent metric (\ref{eq:ADM
  line element - emergent - spherical}), which is associated to the constraint
$\tilde{H}=\tilde{H}^{\rm grav}+\tilde{H}_s^{\rm matter}$ given by the
expressions (\ref{eq:Hamiltonian constraint - modified - non-periodic}) and
(\ref{eq:Hamiltonian constraint - PF - spherical - EMG}) that satisfy the
hypersurface-deformation brackets (\ref{H_x,H_x bracket})--(\ref{eq:H,H bracket
  - emergent}); the dynamics are generated by the constrains $\tilde{H}$ and
$H_x$ with lapse $N$ and shift $N^x$; and the expression (\ref{eq:ADM line element
  - emergent - spherical - conformal modification}) is a covariant tensor not
strictly related to the space-time geometry.

In the latter case, the space-time geometry is given by the emergent metric
(\ref{eq:Conformal metric in conformal frame}), which is associated to the
constraint $\bar{H}$ given by the expression (\ref{eq:Conformal constraint})
that satisfies the hypersurface-deformation brackets (\ref{H_x,H_x bracket -
  conformal})--(\ref{eq:H,H bracket - emergent - conformal}); the dynamics are
generated by the constraints $\bar{H}$ and $H_x$ with lapse $\bar{N}$ and shift
$\bar{N}^x$.  Dynamics are, therefore, different in each case: Setting $N=1$
will generate different equations of motion than $\bar{N}=1$, the latter
corresponding to $N=\omega^{-1}$ in the original frame (using $\tilde{H}$).

The conclusion that both the space-time geometry and the dynamics are
different leads to further questions of which is the correct system for
physical applications.  While such a question can only be answered by experiments or
observations, we may discuss theoretical or conceptual aspects favoring one
over another.  First, the fact that the space-time geometry is changed in the conformal
case leads us to question whether the original matter coupling is still valid.  While
the system remains covariant, one would expect the correct theory to present
the perfect fluid directly coupled to the space-time metric, in this case given
by $\bar{g}_{\mu \nu}$ instead of the original $\tilde{g}_{\mu\nu}$.  If we
amend this, following \cite{EmergentFluid}, the actual perfect fluid contribution to
the Hamiltonian constraint should be given by
\begin{eqnarray}
    \!
    \bar{H}^{\rm matter}_s \!\!\!&=&\!\!\!
    \sqrt{ s P_T^2 + \bar{q}^{x x} (H_x^{\rm matter})^2}
    - \sqrt{\det{\bar{q}}} P_q \left(E^x , \frac{P_X^2}{P_T^2}\right)
    \nonumber\\
    \!\!&&\!\!
    - 4\pi \sqrt{E^x} E^\varphi \bar{\lambda}_0 P_0 \left(E^x , \frac{P_X^2}{P_T^2}\right)
    \,,
    \label{eq:Hamiltonian constraint - PF - spherical - EMG - conformal}
\end{eqnarray}
which differs from $\omega^{-1}\tilde{H}^{\rm matter}_s$, with
$\tilde{H}^{\rm matter}_s$ given by (\ref{eq:Hamiltonian constraint - PF -
  spherical - EMG}), by using $\bar{q}^{x x}$ instead of $\tilde{q}^{x x}$ and
an overall factor of $\omega^{-1}$.  The use of the amended matter constraint
contribution (\ref{eq:Hamiltonian constraint - PF - spherical - EMG -
  conformal}) will, in turn, change the dynamics even further and, as a
consequence, notions of the fluid frame will change, rendering the motivation of certain gauge
choices more difficult.

Finally, we may expect canonical quantization to differ significantly between
the different systems because the corresponding constraint, either $\tilde{H}$
or $\bar{H}$, with its associated brackets must be used in the quantization
procedure.  In this application, it is the original system, with $\tilde{H}$,
that is the more natural choice because it is the simpler constraint: The
overall factor $\omega^{-1}$, required for a nonsingular solution, depends on the complicated phase-space function
$\mathcal{M}$, which must be turned into a composite operator in the quantum
theory, leading to ordering issues among others.

\end{appendix}


\end{document}